\documentclass[conference]{IEEEtran}
\IEEEoverridecommandlockouts
\usepackage{hyperref}
\usepackage{booktabs}

\usepackage{cite}
\usepackage[pdftex]{graphicx}
\DeclareGraphicsExtensions{.pdf,.jpeg,.png}
\usepackage{amsmath}
\usepackage{comment}
\usepackage{algorithmic}
\usepackage{array}
\usepackage{fixltx2e}
\usepackage{url}
\usepackage{color,soul}
\usepackage{caption}
\usepackage{subcaption}
\usepackage{multirow}
\usepackage{authblk}
\usepackage{xcolor}
\usepackage{hyperref}

\begin{document}
\bstctlcite{IEEEexample:BSTcontrol}
\title{WHYPE: A Scale-Out Architecture with\\ Wireless Over-the-Air Majority for\\ Scalable In-memory Hyperdimensional Computing}


\author[$\dagger$*]{Robert Guirado}
\author[$\dagger$]{Abbas Rahimi}
\author[$\dagger$]{Geethan Karunaratne}
\author[*]{Eduard Alarc\'on}
\author[$\dagger$]{Abu Sebastian}
\author[*]{Sergi Abadal}

\affil[$\dagger$]{IBM Research -- Zurich, R\"{u}schlikon, Switzerland}
\affil[*]{Universitat Polit\`ecnica de Catalunya, Barcelona, Spain}

\maketitle

\begin{abstract}
Hyperdimensional computing (HDC) is an emerging computing paradigm that represents, manipulates, and communicates data using long random vectors known as hypervectors. Among different hardware platforms capable of executing HDC algorithms, in-memory computing (IMC) has shown promise as it is very efficient in performing matrix-vector multiplications, which are common in the HDC algebra. Although HDC architectures based on IMC already exist, how to scale them remains a key challenge due to collective communication patterns that these architectures required and that traditional chip-scale networks were not designed for. To cope with this difficulty, we propose a scale-out HDC architecture called WHYPE, which uses wireless in-package communication technology to interconnect a large number of physically distributed IMC cores that either encode hypervectors or perform multiple similarity searches in parallel. In this context, the key enabler of WHYPE is the opportunistic use of the wireless network as a medium for over-the-air computation. WHYPE implements an optimized source coding that allows receivers to calculate the bit-wise majority of multiple hypervectors (a useful operation in HDC) being transmitted concurrently over the wireless channel. By doing so, we achieve a joint broadcast distribution and computation with a performance and efficiency unattainable with wired interconnects, which in turn enables massive parallelization of the architecture.  
Through evaluations at the on-chip network and complete architecture levels, we demonstrate that WHYPE can bundle and distribute hypervectors faster and more efficiently than a hypothetical wired implementation, and that it scales well to tens of receivers. We show that the average error rate of the majority computation is low, such that it has negligible impact on the accuracy of HDC classification tasks.
\end{abstract}

\section{Introduction} \label{introd}
\vspace{-0.1cm}
Hyperdimensional computing (HDC) is an emerging computational framework and is based on the observation that key aspects of human memory, perception and cognition can be explained by the mathematical properties of hyperdimensional spaces comprising high-dimensional vectors known as hypervectors~\cite{hdcintro}. The $d$-dimensional hypervectors are generated using (pseudo)random process such that their components are independent and identically distributed. When the dimensionality ($d$) is in the thousands, a large number of quasi-orthogonal hypervectors exist. This allows HDC to combine existing hypervectors into new hypervectors using well-defined vector operations, such that the resulting hypervector is unique and preserves the dimensionality. To review HDC and related computational models in detail refer to~\cite{HDC_Rev_PI}.  

HDC has been employed in a range of applications including cognitive computing~\cite{PlateAnalogy2000,KanervaDollar2010}, robotics~\cite{NeubertRobotics2019}, distributed computing~\cite{VSA_Workflow,SimpkinScalable2018,TomsettDemonstrationOrch2019}, communications~\cite{CollectiveComm,Dependable_MAC_HD,Kim2018HDM,Hsu2019Collision,Hsu_HDM2,Hersche2021}, and in various aspects of machine learning. See~\cite{HDC_Rev_PII} for a comprehensive review. HDC has achieved particularly notable accuracy in machine learning applications that demand few-shot learning, where other alternative approaches have generally struggled~\cite{InMemFSCIL2022,FSCIL2022,Karunaratne2021RobustHM,MoinWearable2021,RahimiBiosignal2019,oneshot}. Among other advantages, HDC is extremely robust in the presence of failures, defects, variations, and noise, all of which are synonymous to ultra-low energy computation. For instance, it has been shown that HDC degrades gracefully in the presence of various faults in comparison to other alternative classifiers: HDC tolerates intermittent errors~\cite{HDC2016}, permanent hard errors in memory~\cite{Li3DVRRAM2016} or in logic~\cite{WuNanotube2018}, and spatio-temporal variations~\cite{HDC_NatElec20} in emerging memory technologies. In a similar vein, it also tolerates noise and interference in the communication channels~\cite{Kim2018HDM,Hersche2021}. These demonstrate robust operations of HDC under low signal-to-noise ratio and high variability conditions thanks to the special brain-inspired properties of HDC: (pseudo)randomness with i.i.d. components, high-dimensionality, and holographic representations (see~\cite{HDC2016} for more details). 

What different HDC algorithms have in common is that they operate on wide vectors. Therefore, HDC calls for architectures that handle operations on a large number of wide vectors efficiently. One of the key operation of HDC is similarity search. It compares an input hypervector with a typically large number of hypervectors that are stored in an associative memory. As a similarity metric, dot-product is often used. This provides a natural fit to exploit in-memory computing (IMC) for HDC~\cite{HDC_NatElec20,InMemFSCIL2022}. An IMC core departs from the von Neumann architectures which move data from a processing unit to a memory unit and vice versa by exploiting the possibility of performing operations (dot products, in our case) within the memory device itself~\cite{memorydevices}. This improves both time complexity and energy consumption of the architecture.

IMC systems have been proposed recently to execute HDC tasks using hypervectors as wide as 10,000-bit~\cite{HDC_NatElec20}. As further elaborated in Section \ref{sec:bcg}, IMC cores are capable of performing similarity searches through dot-products with unprecedented energy efficiency, i.e. $\sim$100$\times$ more efficiently than a digital accelerator~\cite{HDC_NatElec20}. This has sparked interest in HDC systems that can handle a large search space. For example, certain applications require to continually add new hypervectors for representing novel classes in the incremental learning regime~\cite{FSCIL2022,InMemFSCIL2022}, and performing similarity search on them, that can grow over thousands hypervectors. Yet still, how to scale HDC architectures to perform searches across a large number of classes remains unclear due to the associated challenges.

HDC architectures can be scaled by either increasing the size of the IMC cores to accommodate many hypervectors (scale-up) or by deploying multiple moderately-sized IMC cores to execute the similarity searches in parallel (scale-out). On the one hand, scaling up requires large IMC cores for the architecture to be usable in incremental learning applications. This poses a fundamental problem in terms of array impedances and programming complexity for the IMC core~\cite{ScaleUpXbar}. On the other hand, scaling out implies distributing wide hypervectors across a potentially large number of modules, which puts a large pressure on the system interconnect. More specifically, such an architecture generates reduction and broadcast communication patterns for which conventional Networks-on-Chip (NoC) and Networks-in-Package (NiP) suffer to deliver competitive performance. 

To address the scalability problem of IMC-based HDC architectures, we propose to use wireless communications within the computing package. Wireless Network-on-Chip (WNoC) have shown promise in alleviating the bottlenecks that traditional NoC and NiP face, especially for the collective traffic patterns that appear when scaling out HDC architectures \cite{Laha2015, ahmed2020asymmetric, jog2021one, micro2022, wiplash}. 
To that end, WNoCs provide native broadcast capabilities, 
which are put to use to implement a chip-scale network that, at the same time, is able to bundle multiple hypervectors and distribute the resulting bundled hypervector to a number of physically distributed similarity search engines.

\begin{figure}[!t]
    \centering
    \includegraphics[width=1\columnwidth]{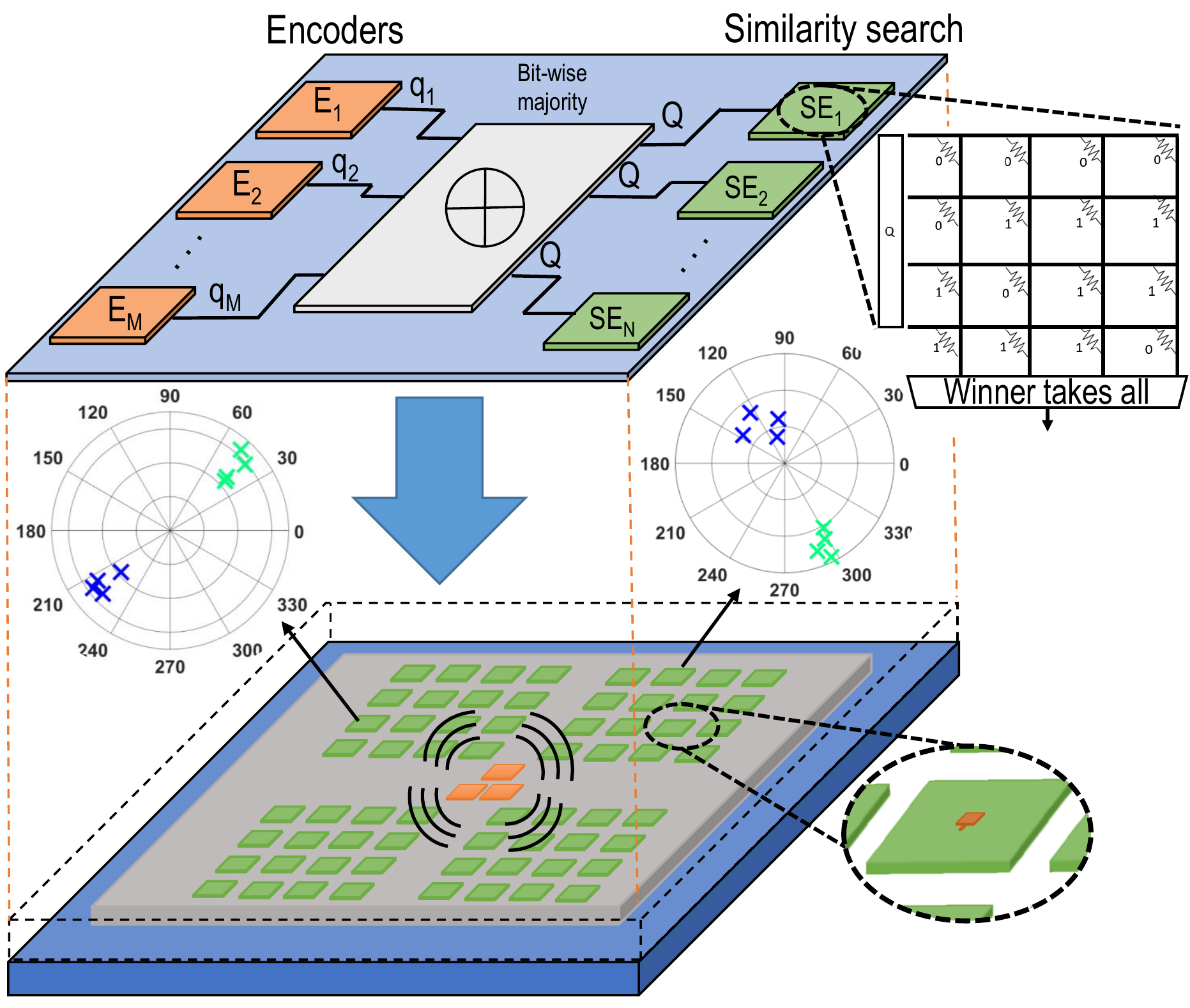}
    \vspace{-0.3cm}
    \caption{Overview of WHYPE, a many-core wireless-enabled IMC platform. Orange encoders map to our wireless transmitters, while green IMCs map to our wireless-augmented IMCs. Bit-wise majority operation required for hypervector bundling is performed via wireless over-the-air computation.}
    \label{fig:large_arch}
    \vspace{-0.5cm}
\end{figure}

In this paper, we present WHYPE, a scale-out architecture that employs wireless over-the-air (OTA) computing to enable scalable in-memory HDC. The architecture, summarized in Fig.~\ref{fig:large_arch}, consists of a set of $M$ encoders that generate query hypervectors. These hypervectors are transmitted simultaneously, bundled over the air, and received by a set of $N$ similarity search engines. Bundling is possible thanks to the OTA computing of the bit-wise majority of the hypervectors, which eliminates the need for a central point of reduction in the architecture. In turn, the concept of OTA is feasible because we have full electromagnetic knowledge of the chip package and we can engineer the constellations to calculate the majority over the air with low error. 

The WNoC proposed for WHYPE is uniquely suited to HDC for two main reasons. On the one hand, it delivers seamless support for the reduction and broadcast patterns required by scale-out HDC architectures. On the other hand, it does so while bypassing the main limitations of wireless interconnects. The generally low aggregate bandwidth is multiplied by $M$ in WHYPE as we allow the concurrent transmission of $M$ hypervectors. Also, the impact of the typically high error rates is minimized thanks to the inherent resilience of HDC algorithms to noise.

In summary, this paper makes the following three novel contributions. First, we present WHYPE, a wireless-enabled architecture for scale-out hyperdimensional computing. Second, we assess the capacity of WHYPE's wireless interconnect to deliver lightweight all-to-all concurrent communications at the chip scale. Third, we evaluate the impact of imperfect wireless communications on the accuracy of the similarity search. It is worth noting that this paper significantly expands on the work presented in \cite{guirado2022wireless} by:
\begin{itemize}
    \item Presenting a complete design for both a wired baseline and the WHYPE architecture, including details on the wireless interfaces and their connection to the encoders and similarity search engines.
    \item Making a comparison of the throughput, area, and power of both the baseline and the WHYPE architecture, which both motivates the need for the proposed architecture.
    \item Extending the performance analysis of WHYPE to the time domain to understand the achievable bandwidth for the OTA computation. 
    \item Evaluating the accuracy of the WHYPE's similarity search considering a distributed dataset.
\end{itemize}

\section{Background}\label{sec:bcg}
To facilitate the understanding of the principles behind WHYPE, here we provide background on the topics of HDC, IMC, and wireless communications at the chip scale.

\subsection{Hyperdimensional Computing}
In this work we focus on a variant of HDC models using 512-dimensional binary hypervectors. Under this setting, by employing a random process, it is easy to find a huge number of non-coincident quasi-orthogonal vectors that exhibit normalized Hamming distance close to 0.5. We call these random hypervectors \textit{atomic} hypervectors. One can further create an \textit{encoder} to operate on these atomic hypervectors by using operations such as binding, bundling (i.e. superposition), and permutation to obtain a composite hypervector describing an object or event of interest. In a classification task, the composite hypervectors, generated from various examples of the same class, can be further bundled together to create a single prototype hypervector representing a class. In this work, the bundling operation is implemented as a logical element-wise majority operation. During training, the prototype hypervectors are stored in the associative memory.

In the inference stage, the query hypervectors of unknown objects/events are generated by the same procedure as in the training stage. The query hypervector is compared to the prototype hypervectors in the associative memory. Then, the chosen label is the one corresponding to the prototype hypervector that has the highest similarity to the query vector. In HDC, the robustness to failure is given by the spreading of information across thousands of dimensions. See~\cite{RahimiBiosignal2019} for more details.

\subsection{In-memory Computing}
With each new technology node, the gap between the speed and efficiency of computation and memory continues to grow. The effects of such a disparity, commonly known as the \textit{memory wall}, have been addressed with novel concepts such as high-bandwidth memory \cite{hbm} or 2.5D and 3D monolithic integration \cite{3d}, among others. However, from an architectural point of view, these solutions are not solving the fundamental bottleneck arising from the need to move large quantities of data from memory and back. Instead, IMC appears as a promising candidate to overcome these challenges \cite{memorydevices}.

IMC is a form of non von-Neumann computing paradigm that leverages the memory unit to perform in-place computational tasks, reducing the amount of data movement and therefore cutting down the latency and energy consumption associated with in-package communication \cite{memorydevices}.  
At the core of IMC is a crossbar array with a memory device lying at each cross point of the array. IMC cores in which these memory devices are resistance-based, and more specifically those based on phase-change memory (PCM) devices, have recently shown promising results~\cite{Y2022khaddamJSSC}. In a resistance-based IMC core, we can program certain values as conductances of cross point memory devices. Executing a matrix-vector multiplication (MVM), essential to any machine learning algorithm, is as simple as, first, tuning conductances to match the matrix values. Second, by exploiting Ohm's law and Kirchhoff's law, inputting the vector as voltages from one side and finally reading the output currents from a perpendicular side.

\begin{figure}[!t]
    \centering
    \includegraphics[width=0.85\columnwidth]{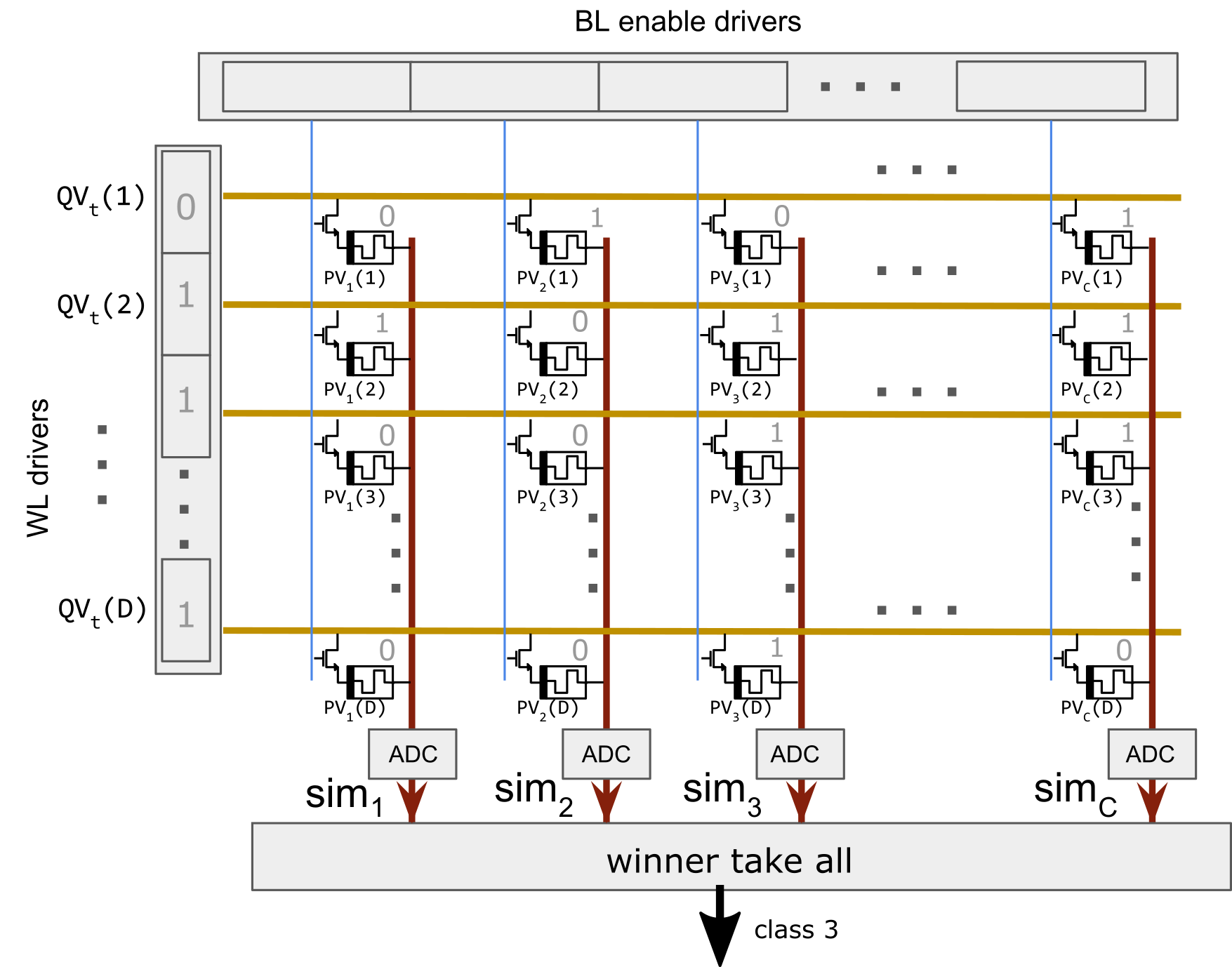}
    \vspace{-0.1cm}
    \caption{Similarity search example in an IMC core. Since the prototype hypervector of the third column is the most similar one to the query vector $Q$, it will output more current than the others and its associated label will be chosen.} 
    \label{fig:dotp}
    \vspace{-0.3cm}
\end{figure}

IMC architectures are capable of executing various HDC operations~\cite{HDC_NatElec20}. In this work, we are particularly interested in the similarity search in the associative memory. As shown in Fig.~\ref{fig:dotp}, since the prototype hypervectors $P_i$ will be programmed in an IMC core, the similarity search through the dot product can be implemented as a MVM with the query hypervector $Q$ as input vector. This allows performing a dot-product in $O(1)$ time complexity.

\subsection{Wireless Network-on-Chip}
Manycore architectures currently rely on NoCs as their interconnect backbone. However, the performance of NoCs can quickly degrade when serving collective communication patterns such as reductions or broadcasts, especially when scaled. In light of this, WNoCs have been recently proposed to complement wired interconnects due to their natural broadcast support, low system-wide latency, and adaptive network topology \cite{ahmed2020asymmetric,adaptive, wiplash, imani2022smart}. Even though WNoC technology is not mature, proof-of-concept designs have been implemented and tested \cite{multichannel}. 

In WNoCs, a core or a cluster of cores are equipped with RF transceivers and antennas \cite{barrier, timoneda,wienna}. This allows them to modulate and transmit data, which propagates through the chip package until being picked up and demodulated by the receivers in the transmitter's range. 
By tuning all antennas to the same channel, WNoC allows to perform low-latency broadcast. However, this comes at the cost of a low aggregate bandwidth as it two simultaneous broadcast transmissions will interfere with each other. Moreover, for the same energy, wireless interconnects are generally much less reliable than wired interconnects due to their \textit{radiative} nature. Fortunately, as described in next sections, the forgiving nature of HDC and the collective nature of its operations minimize the impact of the WNoC disadvantages.

\begin{figure*}[!t]
    \centering
    \begin{subfigure}{0.35\textwidth}
    \includegraphics[width=\textwidth]{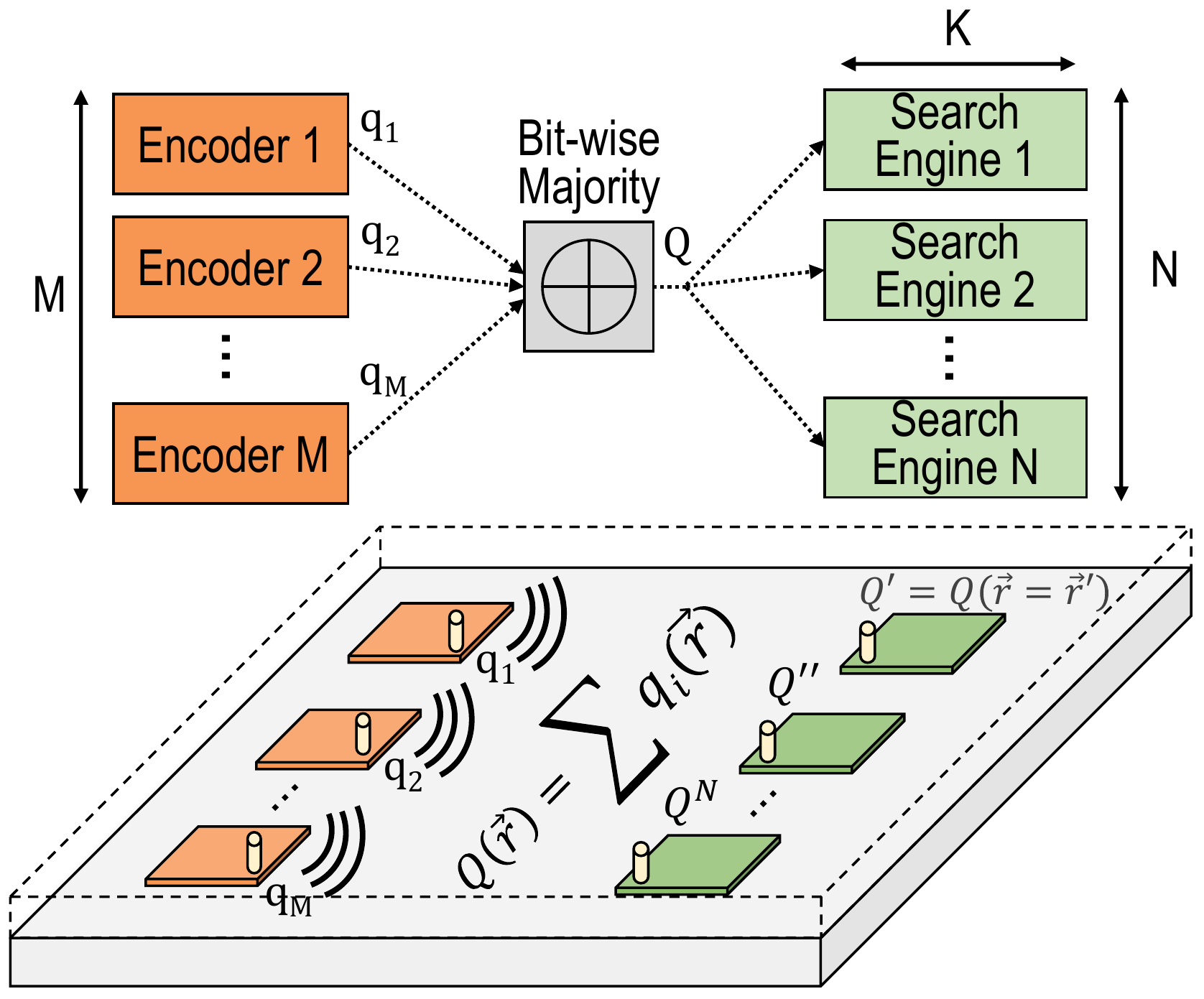}
    \caption{Logical (top) and physical view (bottom).}
    \label{fig:diagram}
    \end{subfigure} \hfill
    \begin{subfigure}{0.63\textwidth} 
    \includegraphics[width=\textwidth]{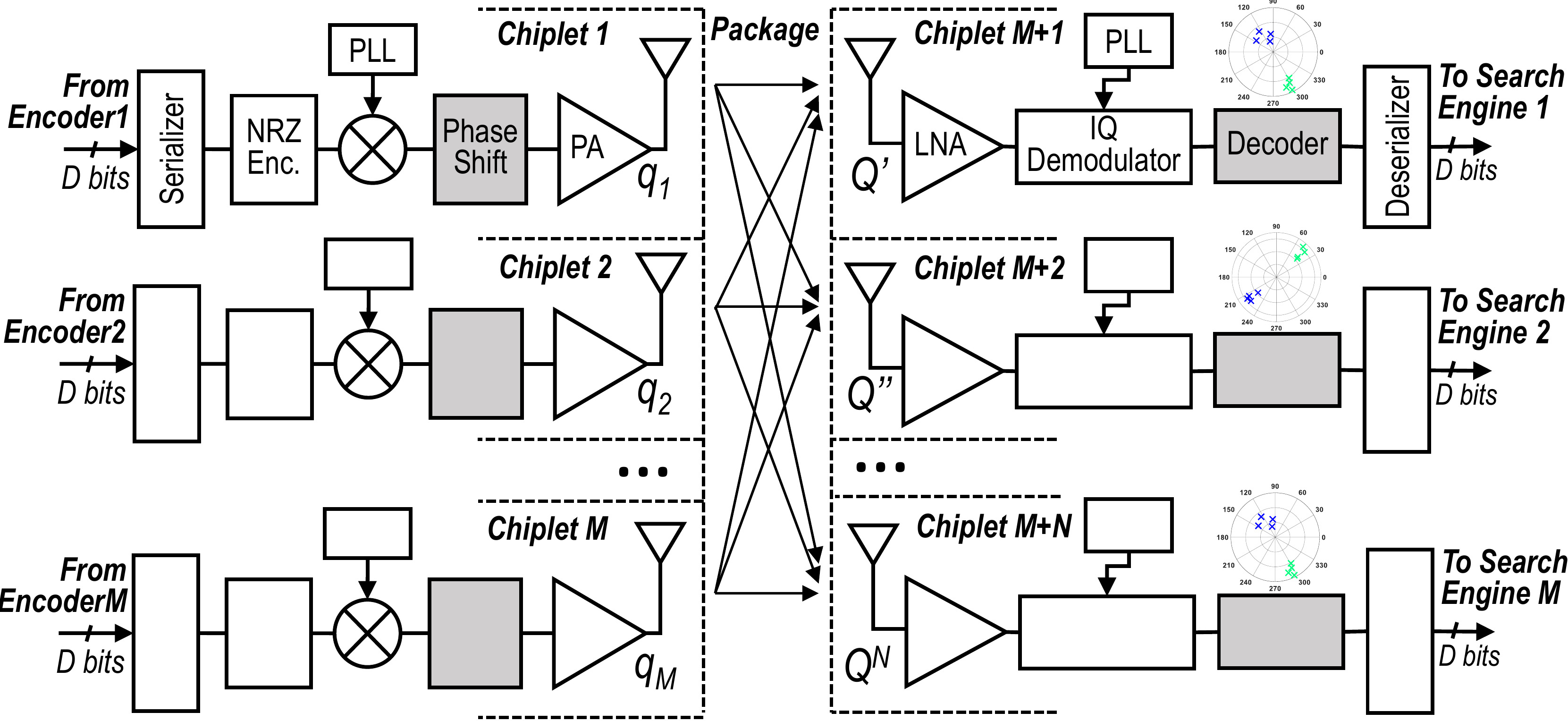}
    \caption{Circuital view (shaded blocks are our contribution).}
     \label{fig:wireless_arch_diagram}
    \end{subfigure}
    \vspace{-0.1cm}
    \caption{Towards a wireless-enabled scale-out HDC architecture with over-the-air computing. (a, top) The architecture involves $M$ encoders generating queries $q_{1}\cdots q_{M}$, the computation of a composite query $Q$ via bit-wise majority, and $N$ IMC cores performing similarity search over multiple copies of $Q$. (a, bottom) In the wireless implementation, the IMC cores receive different versions of $Q$ over space ($Q'\cdots Q^{N}$) that need to be decoded. (b) To enable the decoding of $Q'\cdots Q^{N}$ with low error, $q_{1}\cdots q_{M}$ are modulated with BPSK and shifted so that the overlapped symbols can be decoded easily at the receiver.}
    \vspace{-0.5cm}
\end{figure*}

\section{Motivation} \label{motiv}
The main aim of this work is to propose an architecture that employs IMC cores for HDC and that can be scaled to satisfy the insatiable appetite of the most demanding workloads.
The top chart of Fig.~\ref{fig:diagram} shows a logical diagram of a possible architecture template for an IMC-enabled HDC-based classifier. The encoding system on the left, possibly divided in a number $M$ of parallel encoders, translate the input data into query hypervectors. These hypervectors are then bundled via a bit-wise majority operation, which virtually increases the computation throughput proportionally to the number of bundled vectors. The search engine on the right, possibly composed by $N$ IMC cores storing $K$ class templates each, compares the bundled query hypervectors with the $N\times K$ prototype hypervectors. We note that both the encoders and the search engines can be implemented with IMC cores.

To scale such an architecture, two broad decisions need to be taken from an architectural standpoint. First, whether to scale by increasing the size of the IMC cores (scale-up; increasing $K$) or by placing more similarity search engines in the system (scale-out, increasing $N$). Second, whether the scaling is done in a fully integrated way, i.e. placing all encoders and search engines within a single chip, or using disintegrated alternatives such as the recent chiplet paradigm. Our proposed architecture, called WHYPE, is build upon three main observations:

\begin{figure}[!t]
    \centering
    \includegraphics[width=0.7\columnwidth]{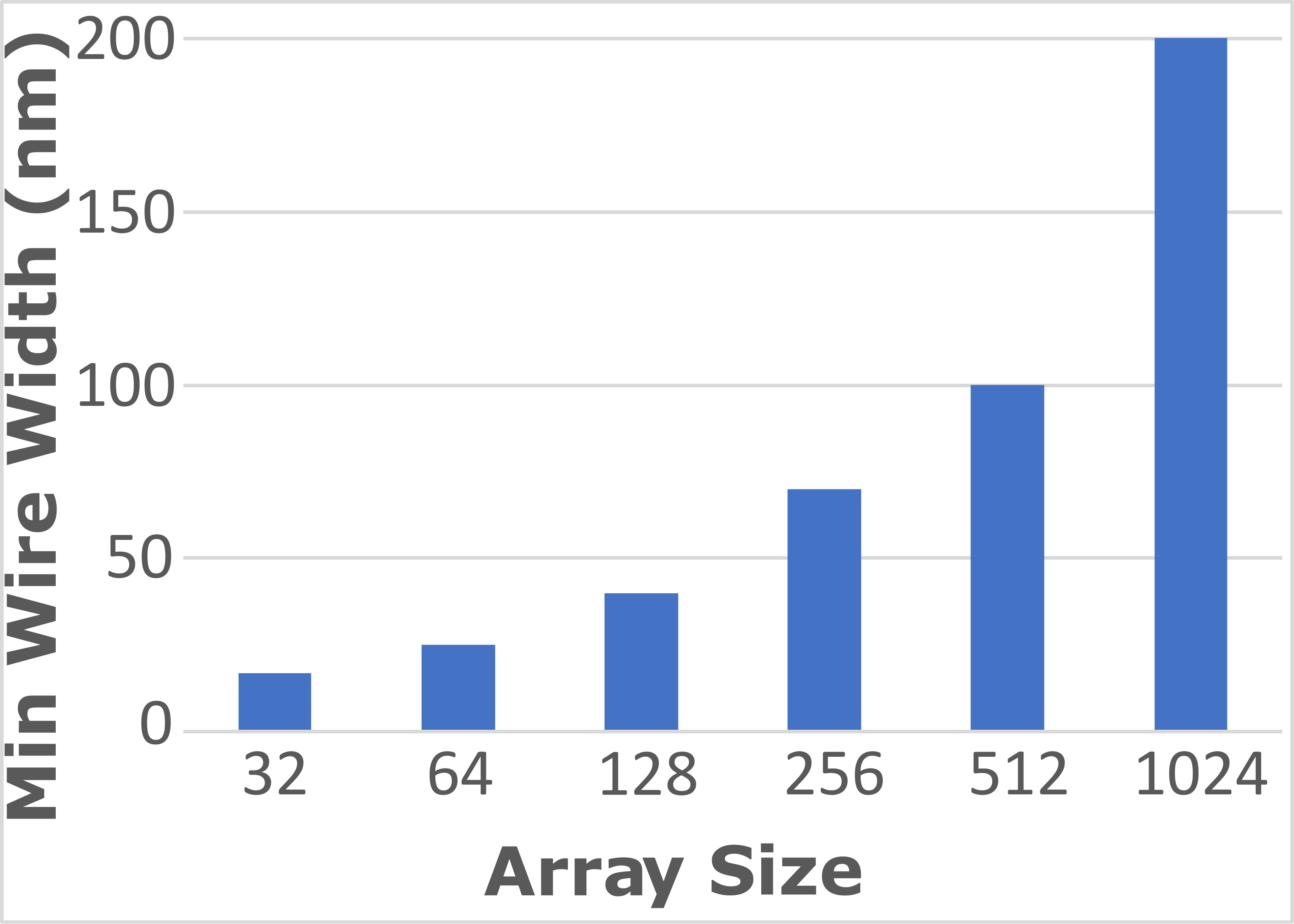}
    \vspace{-0.1cm}
    \caption{Minimum wire width requirement of crossbar arrays as a function of the array size, in order to maintain the same effect from IR drop and other non-idealities on the MVM accuracy}
    \label{fig:scaleIMC}
    \vspace{-0.5cm}
\end{figure}

\vspace{0.1cm}\noindent
\textbf{Observation 1: In-memory cores do not scale up well.} Increasing the size $K$ of the IMC arrays allows accommodating a larger number and of prototype hypervectors (classes). However increasing the crossbar size brings with itself a number of non-ideal properties. For instance, IR drop across the array is increased due to the interconnect resistance of longer wires and RC latency is increased due to the increased parasitic capacitance. To counter these problems, one has to sacrifice one or few important metrics related to performance, power and area. For e.g. the IR drop can be reduced by increasing the wire width. The required minimum wire width at different array sizes to maintain the same effect of IR drop is given in Fig.~\ref{fig:scaleIMC} and in \cite{scaleupIMC}. As shown in the figure, the wire width exponentially rises, limiting the scalability of array sizes. Moreover, the complexity of weight programming also increases with the array size~\cite{ScaleUpXbar}. 
There are already IMC prototypes supporting up to 256 prototype hypervectors~\cite{InMemFSCIL2022}. This may be stretched up to 512 and 1024 but most likely not to bigger sizes due to the rise of non-ideal properties.
\textit{Therefore, scale-out architectures that employ multiple, but relatively small IMC cores, may be preferable in this scenario.}

\vspace{0.1cm}\noindent
\textbf{Observation 2: There exist different application spaces in terms of required search throughput, and the number of inputs and classes.} 
Following ~\cite{RahimiBiosignal2019}, a generic scalable HDC architecture for various workloads should include a set of encoders (shown in the left hand-side of Fig.~\ref{fig:large_arch}) and a set of similarity search engines (shown in the right hand-side of Fig.~\ref{fig:large_arch}). The number of encoders is typically determined by the demand of application, ranging from few encoders to operate on data from different sensory modalities~\cite{ChangEmotion2019,MitrokhinCNN2020}, to a larger number of encoders working with independent streaming channels~\cite{Hersche2021}. Similarly, the number of required similarity search engines is determined by the application.
It is based on the number of classes that the application has to support which ranges e.g., from a handful of classes~\cite{HDC_NatElec20,InMemFSCIL2022} to over one thousand classes~\cite{FSCIL2022}, or half million classes~\cite{ExteremeLearning_NIPS21}. Further, the number of \emph{active} classes could change over time, e.g., in a continual learning regime at the beginning there are about 60 classes that can grow up to 1600 during the course of incremental learning~\cite{FSCIL2022,InMemFSCIL2022}.
In computational terms, this means that the capacity of the encoders and similarity search engines is highly dependent on the application. We therefore aim for a generic architecture to be able to scale out by many inputs or classes coming in. Assuming a fully integrated design, architects would need to carefully dimension the encoders and similarity search engines in order to cater to the needs of a specific application. This fundamentally limits the reuse of the system in other domains, which may either require faster/broader or smaller/more efficient searches and where the designed architecture would either underperform or waste area and power. \textit{Such an observation suggests that a disintegrated architecture, possibly chiplet-based, would be a viable route for scaling-out IMC-based HDC architectures.} 

\vspace{0.1cm}\noindent 
\textbf{Observation 3: Scaling out HDC architectures is costly.}
In particular, scaling the architecture template described in Fig.~\ref{fig:diagram} quickly leads to a communication bottleneck, especially in chiplet-based systems. First, even though bundling compresses $M$ hypervectors of length $d$ into a single composite query, that comes at the cost of a heavy reduction communication flow. This $M$-to-1 traffic pattern is due to the need to bring the different hypervectors to a common circuit performing the bit-wise majority operation. The majority operation, besides the significantly large buffers required to store the $M$ inputs and the output hypervector, takes $M$ inputs of size $d$ with a known complexity of $d \times M^{3.5}$ gates \cite{choudhary2019generalized}. Only when the bundling is finalized, the output can be distributed to the $N$ IMC cores. However, broadcast patterns are generally expensive with growing $N$ \cite{ahmed2020asymmetric}. 

In summary, bundling and data movement quickly becomes a bottleneck when scaling, especially in disintegrated architectures. In chiplet-based systems, significant energy ($\sim$1 pJ) and latency ($\sim$20 ns) can be expected per hop due to connectivity and I/O pin limitations \cite{simba}, whereas hop counts will grow at least proportionally to the number of encoders $M$ and the number of search engines $N$ due to the reduction and broadcast flows, respectively \cite{wienna}. \textit{Hence, alternative implementations of the reduction-majority-broadcast pattern are necessary to enable the effective scale-out of HDC architectures.}

\section{WHYPE: A Wireless-Enabled Architecture for Scale-Out of Hyperdimensional Computing}
As depicted in Section \ref{motiv}, even though disintegrated scale-out is a desirable choice for scaling HDC systems, implementing such architectures with wired interconnects is challenging. This is because of three main reasons:
\begin{itemize}
    \item The bundling operation generates a reduction pattern that can create a communication bottleneck at the vicinity of the bundling circuit.
    \item The query distribution requires broadcast communication, which is inherently costly for wired interconnects in general and in chiplet-based systems in particular.
    \item The bundling operation creates an implicit barrier that forces encoding and similarity search to be done sequentially. While all three operations can be pipelined, the end-to-end latency increases.
\end{itemize}

Our proposed architecture, called WHYPE, addresses the three problems of wired scale-out at once. To that end, we augment a many-core HDC system with a wireless chip-scale network specifically designed to eliminate the need to transfer all the hypervectors to a central point for bundling. Fig.~\ref{fig:wireless_arch_diagram} shows the proposed implementation of WHYPE, which is composed by $M$ encoders augmented with wireless transmitters alongside $N$ IMC cores augmented with wireless receivers. The transmitters and receivers are slightly modified versions of a simple BPSK modulator and a coherent receiver and decoder, respectively. The frequency of operation is high, e.g. 60 GHz, to minimize the area and power overhead of the RF circuits and the antenna. Finally, given the monolithic nature of the system, we assume that the clocks of all transmitters are synchronized.

The mode of operation is as follows. All the encoders broadcast, simultaneously and through the same wireless channel, the queries $q_{1}\cdots q_{M}$ that must be bundled to form $Q$. As a consequence of wave propagation within the package, the receivers will obtain $N$ different versions of the superposition of the $M$ symbols transmitted concurrently, $Q'\cdots Q^{N}$. Each receiver will then take its received signal and attempt to decode a single bit representing the majority operation of the $M$ transmitted symbols. In other words, the decoders will be carefully designed to achieve $Q'=Q$, $ \cdots$, $Q^{N}=Q$ with high probability so that, effectively, \textit{the bit-wise majority of the transmitted hypervectors is computed over the air.}

\begin{figure}[!b]
    \centering
    \includegraphics[width=1\columnwidth]{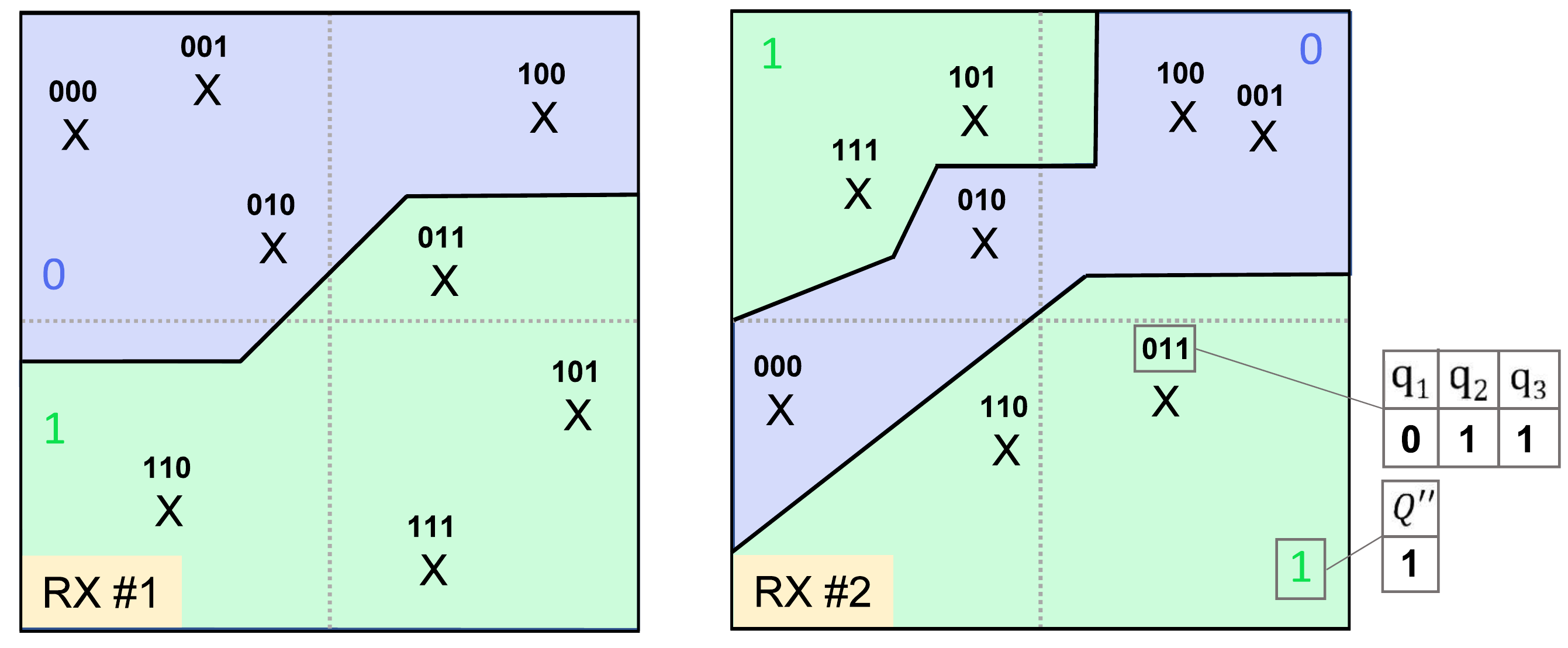}
    \vspace{-0.5cm}
    \caption{Example of decision regions of over-the-air (OTA) majority computation for three transmitters $\{q_1, q_2, q_3\}$ at two distinct receivers. Blue/green regions map to 0/1.}
    \label{fig:regions_ex}
\end{figure}

To further illustrate the idea behind over-the-air majority computation, Fig.~\ref{fig:regions_ex} shows a sample constellation that could result from the superposition of three transmissions. Each point represents one of the $2^{3}=8$ possible combinations. Since we are not interested in the values of the three transmitted bits, but rather in the result of the majority operation, the decoders only need to distinguish between the combinations that lead to $maj(\cdot) = 0$ and $maj(\cdot) = 1$, respectively. The key objective, then, is to modulate the information in each transmitter so that the received constellations form two easily separable clusters for $maj(\cdot) = 0$ and $maj(\cdot) = 1$. WHYPE achieves this with a very simple variant of source coding, i.e. through pre-assigned phase shifts. As detailed further in Section~\ref{sec:method1}, an exhaustive search is performed offline to find the transmitters' phase shifts leading to easy-to-decode majority at the receivers.

At its core, WHYPE exploits three key opportunities, which allow to boost the value and minimize the disadvantages of wireless chip-scale communications:

\vspace{0.1cm} \noindent
\textbf{Key Opportunity 1: Over-the-air computing is possible because the channel is static and known beforehand.} OTA computing is certainly not new, but it has generally been hindered by the need of accurate and up-to-date channel state information, which is extremely hard to guarantee in conventional wireless networks \cite{altun2022magic}. In contrast, the in-package scenario is static and allows for a pre-characterization of the channel \cite{Matolak2013CHANNEL}, hence allowing for an OTA calculation of the majority operations required by the bundling of hypervectors.

\vspace{0.1cm} \noindent
\textbf{Key Opportunity 2: The inherent broadcast nature of wireless communication allows for a single-hop distribution of bundled hypervectors.} By using omnidirectional antennas such as vertical monopoles \cite{Pano2020a}, data is naturally broadcast with a latency and efficiency hard to achieve with wired on-chip networks \cite{orthonoc}. This feature, together with the OTA bundling, allows to eliminate the communication bottleneck of HDC scale-out architectures.

\begin{figure}[!h]
    \centering
    \includegraphics[width=0.8\columnwidth]{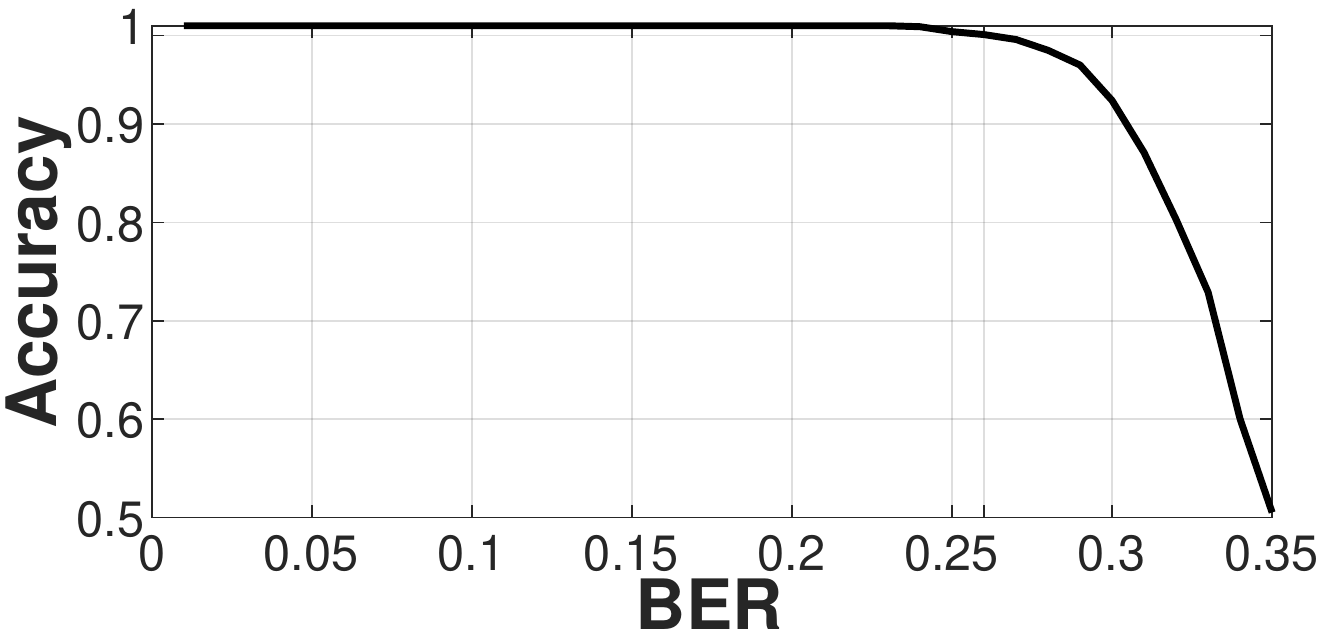}
    \vspace{-0.2cm}
    \caption{Impact of the bit error rate (rate of erroneous bits in a bundled hypervector) on the accuracy of a classification task under the conditions described in Section \ref{sec:searchMethod}.}
    \label{fig:ber_acc} 
    \vspace{-0.3cm}
\end{figure}

\vspace{0.1cm} \noindent
\textbf{Key Opportunity 3: The resilience of the HDC paradigm to errors makes it tolerant to unreliable communication}. A drawback of wireless communications in general (and of OTA computing in particular) is that it can suffer from relatively high error rates when compared to wired communications. This generally leads to low energy efficiency. 
However, as we illustrate in Fig.~\ref{fig:ber_acc}, HDC is inherently resistant to errors and opens the door to the use of unreliable wireless communications without compromising the energy efficiency and scalability of the architecture.

\begin{figure*}[!t]
    \centering
    \includegraphics[width=0.53\textwidth]{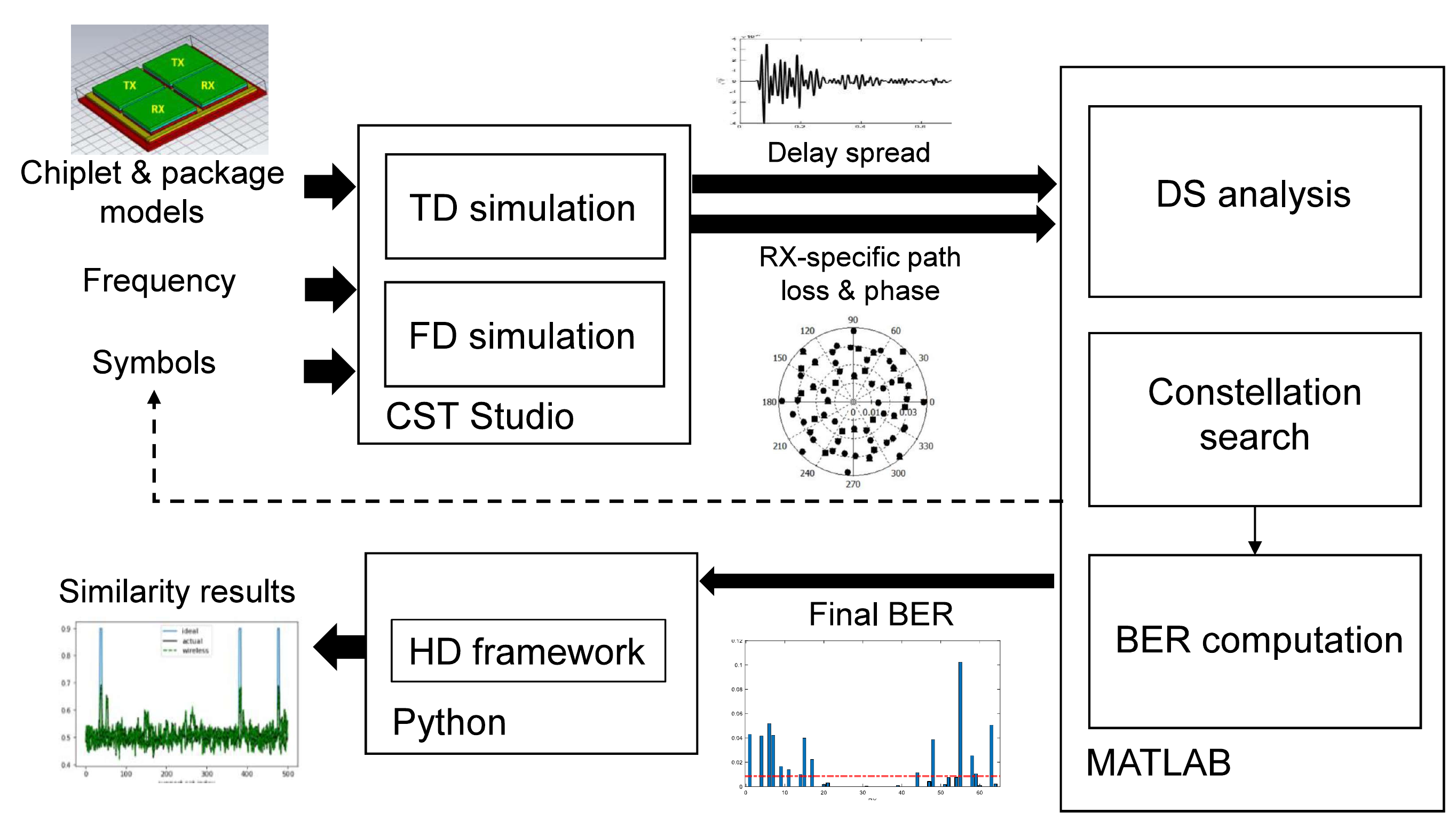}\hfill
    \includegraphics[width=0.47\textwidth]{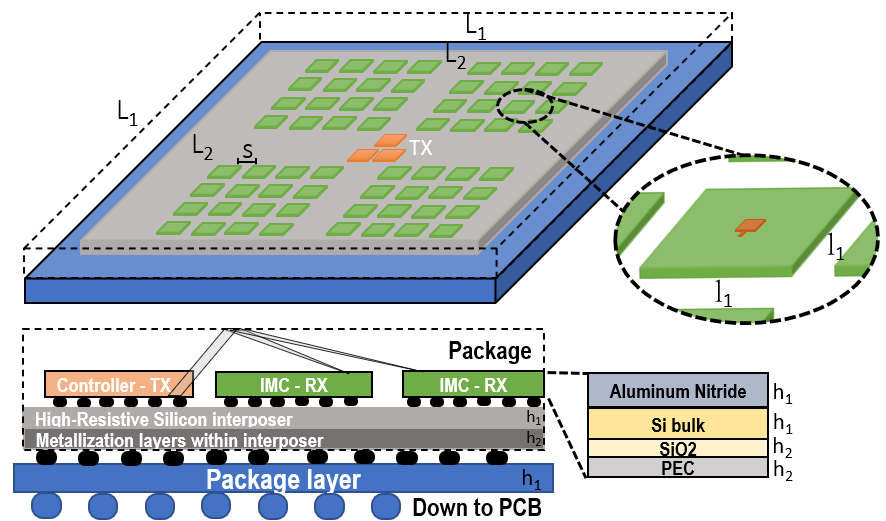}
    \vspace{-0.4cm}
    \caption{Overview of the evaluation methodology and layout of a sample architecture with 3 TXs and 64 RXs. The package is enclosed in a metallic lid and empty spaces are filled with vacuum. $h_1=0.1$ mm; $h_2 = 0.01$ mm; $l_1 = 7.5$ mm; $s = 3.75$ mm; $L_1 = 33$ mm; $L_2 = 30$ mm.}
    \label{fig:meth}\label{fig:arch}
    \vspace{-0.5cm}
\end{figure*}

\section{Methodology} \label{meth}
Fig. \ref{fig:meth} summarizes the procedures followed to evaluate WHYPE from the perspectives of wireless communications and HDC architectures. First, the computing package has been modeled in CST \cite{cst} assuming a disintegrated architecture with $M+N$ chiplets with their respective antennas and transceivers, implementing the scheme from Fig.~\ref{fig:diagram}. The output of CST simulations is fed to MATLAB to assess the bit error rate (BER) of the resulting constellations. The BER is then used in a python-based HDC framework to characterize the impact of imperfect communication on the HDC classification accuracy. 

Next, we describe the methods to obtain the source coding in Sec.~\ref{sec:method1}, to assess the speed and reliability of OTA computing in Sec.~\ref{sec:method2}, to evaluate the classification accuracy in Sec.~\ref{sec:method3}, and the details of the classification benchmark in Sec.~\ref{sec:method4}. 

\subsection{Source Coding Optimization}
\label{sec:method1}
As shown in Fig.~\ref{fig:wireless_arch_diagram}, transmitters encode the bits of their queries using a BPSK encoder plus two specific phase shifts (for the symbols '0' and '1', respectively). That is, all symbols will have the same amplitude, but a different phase. Since we let the $M$ encoders to transmit simultaneously, each of the $N$ receivers will observe a slightly different superposition of all the transmitted symbols. This leads to a $N$ different constellations of $2^M$ points each.

In this context, the objective is to select the phase shifts at each transmitter so that all the received constellations are clustered in two separable decision regions, each corresponding to the case where $maj(\cdot)=0$ and $maj(\cdot)=1$, respectively. This optimization process has two constraints. On the one hand, we have to make sure that each transmitter only uses two phases, for its symbol '0' and '1', respectively. On the other hand, the phases at one transmitter have an impact on the constellations of all receivers, which implies that \textit{a joint optimization considering all RXs is needed}.

\begin{figure}[!t]
    \centering
    \includegraphics[width=\columnwidth]{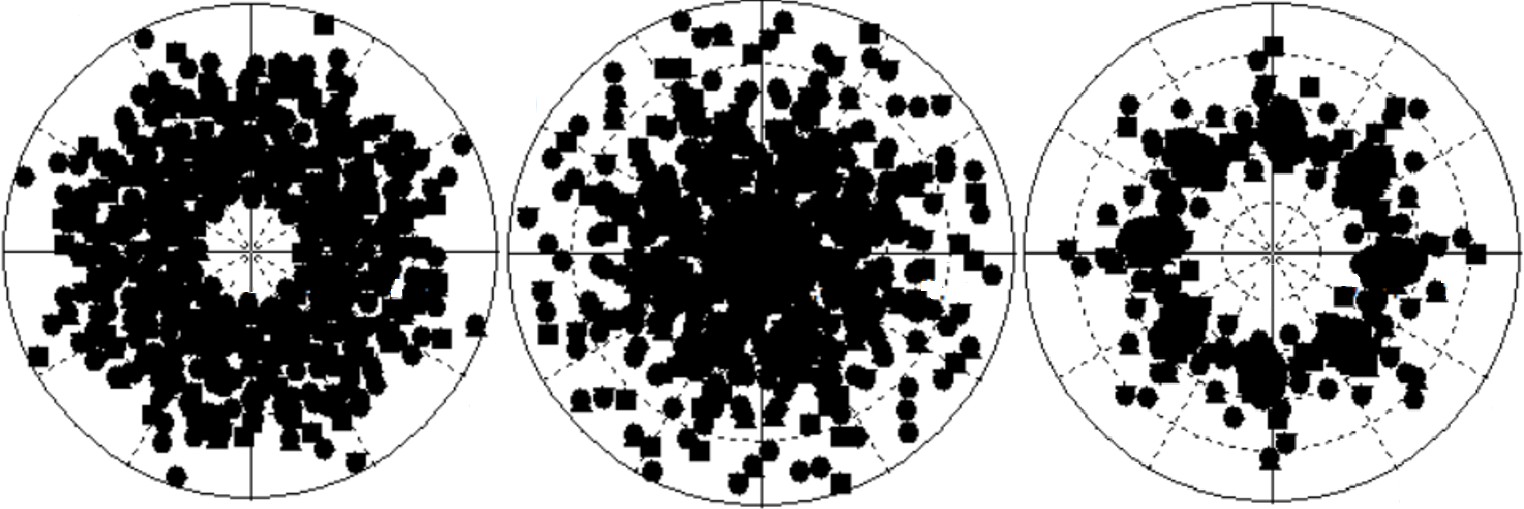}
    \includegraphics[width=\columnwidth]{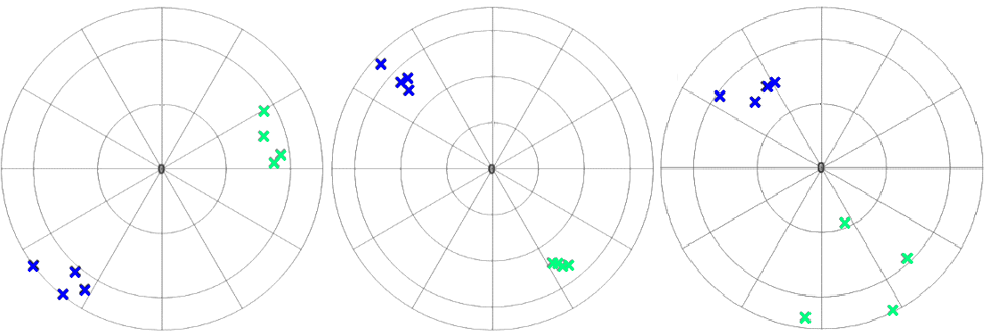}
    \vspace{-0.1cm}
    \caption{Sweep of all possible phase combinations (top) and the one that  minimizes the error rate of the majority computation (bottom). Blue/green symbols map to logical 0/1.}
    \label{fig:ota_constellations}\label{fig:cst_res}
\end{figure}

\begin{figure}[!t]
    \centering
    \vspace{-0.3cm}
    \includegraphics[width=0.8\columnwidth]{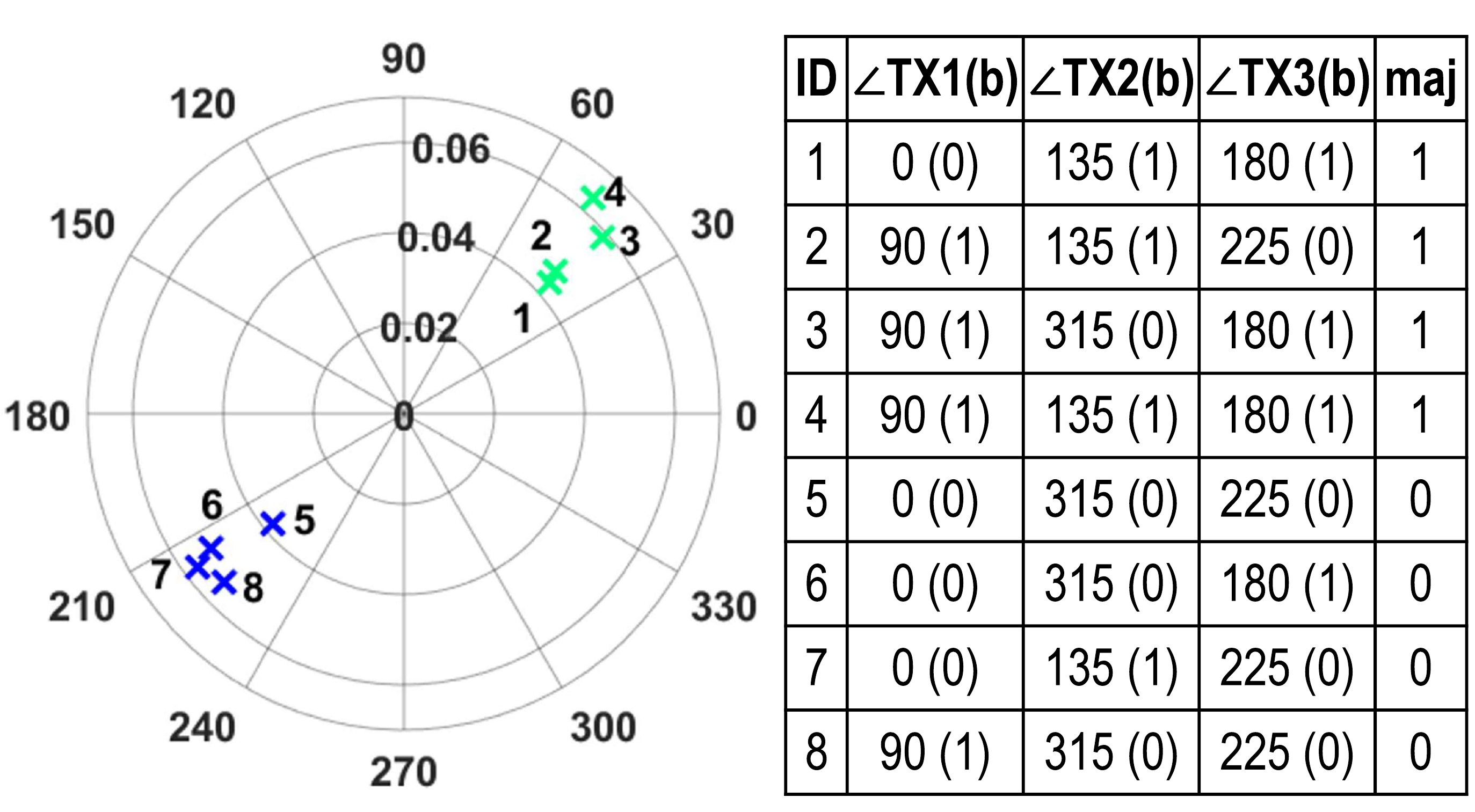}
    \vspace{-0.2cm}
    \caption{Constellation and truth table with transmitted phases/ bits for a specific RX. Blue/green symbols map to logical 0/1.}
    \label{fig:otaasig_tab2} 
    \vspace{-0.4cm}
\end{figure}

The optimization process starts by simulating the simultaneous transmissions in CST. We consider a set of 8 phases in each transmitter (i.e. in 45 degree steps) and evaluate the amplitude and phase obtained at each of the receivers. Then, the results are fed to MATLAB, where the decision regions are computed using the $K$-means clustering algorithm with $K=2$ over the constellation. These regions are used to evaluate the average error across all receivers using the methods described in Sec.~\ref{sec:method2}. The combination of phases leading to the lower average error rate is selected. An illustrative example with three transmitters and three receivers is shown in Fig.~\ref{fig:cst_res}, with the detail of a specific receiver in Fig.~\ref{fig:otaasig_tab2}.

\subsection{Over-The-Air Computing Evaluation}
\label{sec:method2}
\noindent
\textbf{Frequency-Domain Simulation.} As described in the previous section, electromagnetic simulations are needed to assess the performance of the over-the-air computation process. We model an interposer-based package with $M+N$ chiplets, each with its own antenna, with the dimensions depicted in Fig.~\ref{fig:arch}. We sweep the phases in each transmitter and obtain the amplitude and phase at the receivers in the frequency domain. The operating frequency is 60 GHz and the transmission power is 0 dBm per antenna, compatible with existing WNoCs \cite{timoneda, multichannel}. Further, we assume $M=3$ transmitters and $N=64$ receivers, unless otherwise noted. Nevertheless, the analysis can be extended to higher frequencies, different power levels, different package configurations, or different number of transmitters and receivers.

\vspace{0.1cm} \noindent
\textbf{Reliability Analysis.} Once the phases are swept and the candidate constellations are obtained, we compute the BER at each RX, for all the different possible constellations, and choose the one that leads to the lowest average BER across RXs. In all cases, the BER is evaluated approximating the centroids of each cluster as received symbols, and using the analytical expression of error rate of BPSK, which is the modulation used in WHYPE. This yields
\begin{equation}
    \label{eq:bpsk}
    BER^{BPSK} = 0.5\cdot \text{erfc}\bigg({\frac{0.5\cdot d_c}{\sqrt{N_0}}}\bigg),
\end{equation}
where $erfc(\cdot)$ is the complementary error function, $d_c$ is the distance among centroids and $N_0$ is the noise spectral density. 

\vspace{0.1cm} \noindent
\textbf{Transmission Speed Analysis.} To assess the speed at which the information can be modulated reliably, time-domain simulations are required. In particular, we use CST to obtain the channel impulse response $h_{i}(t)$ with simultaneous source excitation from the $M$ transmitters to the receiver $i \in [1, N]$. 
Then, the power-delay profile (PDP) of receiver $i$, $P_{i}(\tau)$, is evaluated as $P_{i}(t) = |h(t)|^2$. The delay spread $\tau^{(i)}_{rms}$ between the sources and receiver $i$ is calculated as the second moment of the PDP as 
\begin{equation}
    \label{eq:ds2}
    \tau^{(i)}_{rms} = \sqrt{\frac{\int (\tau - \bar{\tau_{i}})^2 P_{i}(\tau)d\tau}{\int P_{i}(\tau)d\tau}} ,
\end{equation}
where $\bar{\tau_{i}} = \frac{\int \tau P_{i}(\tau)d\tau}{\int P_{i}(\tau)d\tau}$ is the mean delay of the PDP for receiver $i$. Since all transmitters need to be synchronized, the modulation speed needs to be the same for all transmitters and below the coherence bandwidth $B_{c}$ of the system. Hence, we calculate the coherence bandwidth via the worst-case delay spread among all receivers, $\tau_{rms}$, so that
\begin{equation}
\tau_{rms} = \max_{i} \tau^{(i)}_{rms} \Longrightarrow B_{c} = \frac{1}{\tau_{rms}}.
\end{equation}

\subsection{Similarity Search Evaluation}
\label{sec:searchMethod}\label{sec:method3}
Once the final transmitter phases leading to the lowest average BER are chosen, an in-house python framework is used to evaluate the impact of imperfect OTA majority computation on the accuracy of the classification task. Every similarity search engine connected to a receiver stores 64 different prototype hypervectors, i.e., 64 different classes, each with 512-bit that suffices for the scenario considered in this paper. This is compatible with current experimentally validated IMC cores \cite{HermesVLSI,Y2022khaddamJSSC}. The area and energy evaluations in this paper assumed a scaled version of the IMC from \cite{HermesVLSI,Y2022khaddamJSSC}. Finally, errors coming from the OTA computations are modeled as uncorrelated bit flips over the query hypervectors.

\vspace{0.1cm} \noindent
\textbf{Bundling alternatives.} While the baseline bundling consists of simply computing the bit-wise logical majority result across the different TX bits, we also consider permuted bundling. This bundling consists of permuting the queries in the TXs prior to applying the majority operation on them. By permuting the hypervectors we obtain two benefits. First, this allows the identification of the transmitter of the detected class from the composite query. The second direct benefit of permuting the hypervectors is that it helps increasing the quasi-orthogonality between them. This has a direct impact on accuracy, since the majority operation over multiple non-orthogonal hypervectors (i.e. not permuted) would yield a bundled hypervector that is hard to classify.

\subsection{Classification Benchmarks}
\label{sec:method4}
We perform experiments using the Omniglot dataset \cite{Omniglot_dataset}. The dataset provides handwritten images of characters from 50 different alphabets. The number of characters in each alphabet varies from 14 to 55. In total there are 1623 character classes and 20 example images from each class. The dataset is further divided into a training and test set containing 964 and 659 character classes respectively. The goal of this benchmark is to train an encoder on the image and label data given in the training set and evaluate classification accuracy on the test images. We perform two types of experiments.

\vspace{0.1cm} \noindent
\textbf{Experiment 1: Few-shot Learning.} First, we evaluate the few-shot learning capability of the system. We do this by, first, meta-training an encoder on the training set data and evaluating the few-shot classification accuracy over a series of 1000 episodes. In each episode, we select 100 classes and 20 encoded hypervector examples (shots) from each class within the test set. These support vectors are distributed among the IMC modules on the RXs. From the remaining images of the same 100 classes, we select 1 query image to encode per TX. The encoded query hypervectors are over-the-air bundled (with or without permutation) and received at each RX to perform the similarity search. The final classification accuracy is the average accuracy across the 1000 episodes.

\vspace{0.1cm} \noindent
\textbf{Experiment 2: Continual Learning.} Secondly, we focus on the continual learning capability of the system. For this, we have a similar setting to that of Experiment 1, with few changes. Here, instead of always choosing a fixed 100 classes per episode, we start with 64 classes and gradually incorporate new classes over a series of sessions. In each session, 64 additional classes are selected from the set of unselected classes so far and 5 selected support examples from these novel classes are provided to the IMCs. During a session, queries can be selected from all classes (both novel and old) currently under selection and one each is assigned to the encoders at the transmitters.

\section{Performance Evaluation} 
\label{results}
We next present the evaluation of WHYPE, first from the perspective of the OTA majority in Section \ref{sec:OTAeval}, and then from the perspective of the classification task in Section \ref{sec:HDCeval}. Additionally, we evaluate the area and power overhead of the architecture in Section \ref{sec:OVHDeval}.

\subsection{Over-the-Air Computing}
\label{sec:OTAeval}
Fig.~\ref{fig:cst_res} illustrates the exhaustive search performed in our system with $M=3$ transmitters, shown only in three receivers chosen at random. The bottom chart of the same figure shows the resulting constellations after the optimization. The constellation in another random receiver is shown in more detail in Fig.~\ref{fig:otaasig_tab2}, together with the phases chosen for the three transmitters: 0\textsuperscript{o}/90\textsuperscript{o}, 315\textsuperscript{o}/135\textsuperscript{o}, and 225\textsuperscript{o}/180\textsuperscript{o} for the symbols '0'/'1' of transmitters TX1, TX2 and TX3, respectively. 

\begin{figure}[!t]
    \centering
    \vspace{-0.1cm}
    \includegraphics[width=0.9\columnwidth]{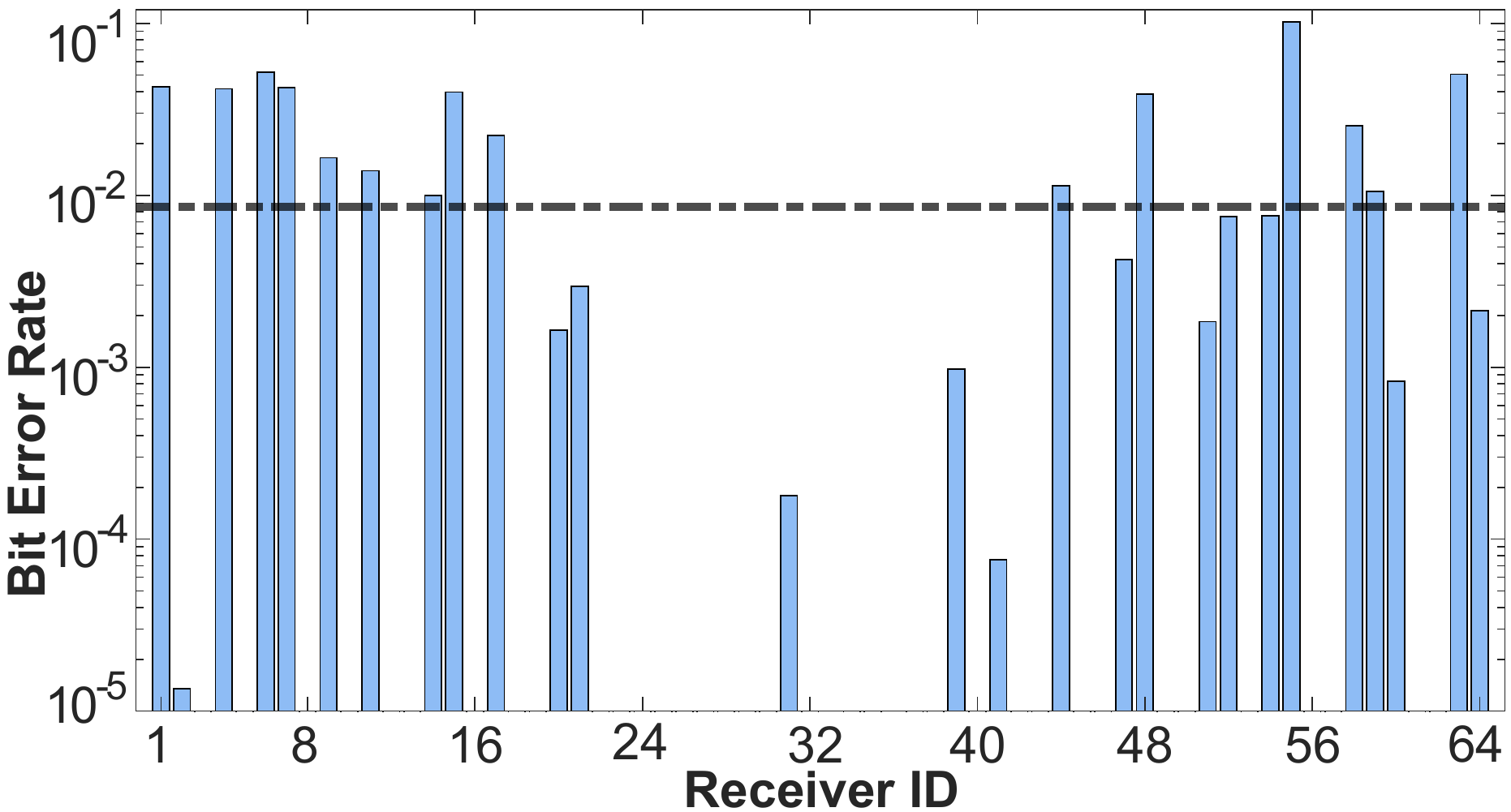}
    \caption{Error rate for each individual RX in the architecture. The dashed line indicates the average value. }
    \label{fig:ber_Rxs} 
    \vspace{-0.3cm}
\end{figure}

\vspace{0.1cm} \noindent
\textbf{Reliability Analysis.} Once the phases are set, we evaluate the error rate of all 64 receivers using Eq.~\eqref{eq:bpsk}. Fig.~\ref{fig:ber_Rxs} shows the BER of each receiver. It can be observed how the BER values are highly dependent on the receiver position, with values as large as $\sim$0.1 and also lower than 10\textsuperscript{-5} in a significant amount of cases. In average, the error rate is below 0.01.

\begin{figure}[!t]
    \centering
    \vspace{-0.2cm}
    \includegraphics[width=0.9\columnwidth]{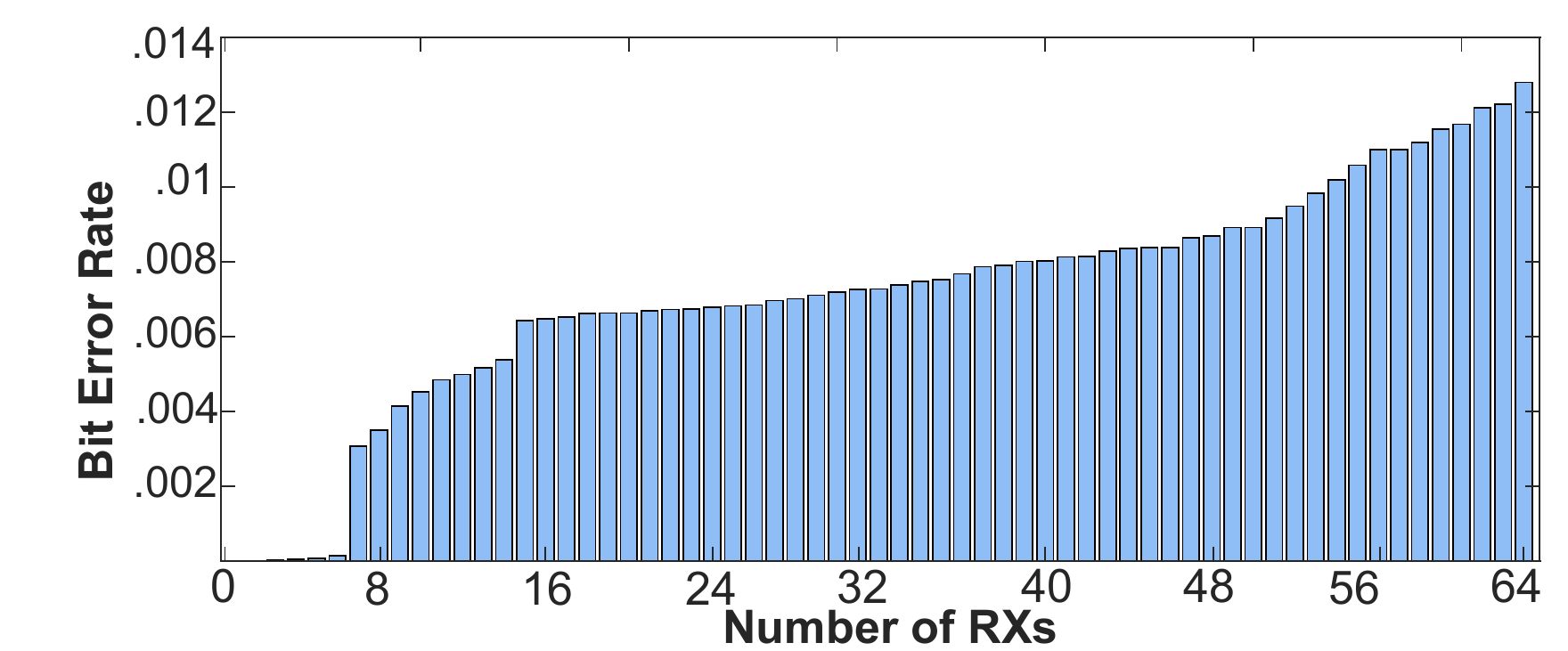}
    \caption{Error rate as a function of the number of receivers.}
    \label{fig:arch_scal} 
    \vspace{-0.4cm}
\end{figure}

\begin{figure}[!t]
    \centering
    \includegraphics[width=0.8\columnwidth]{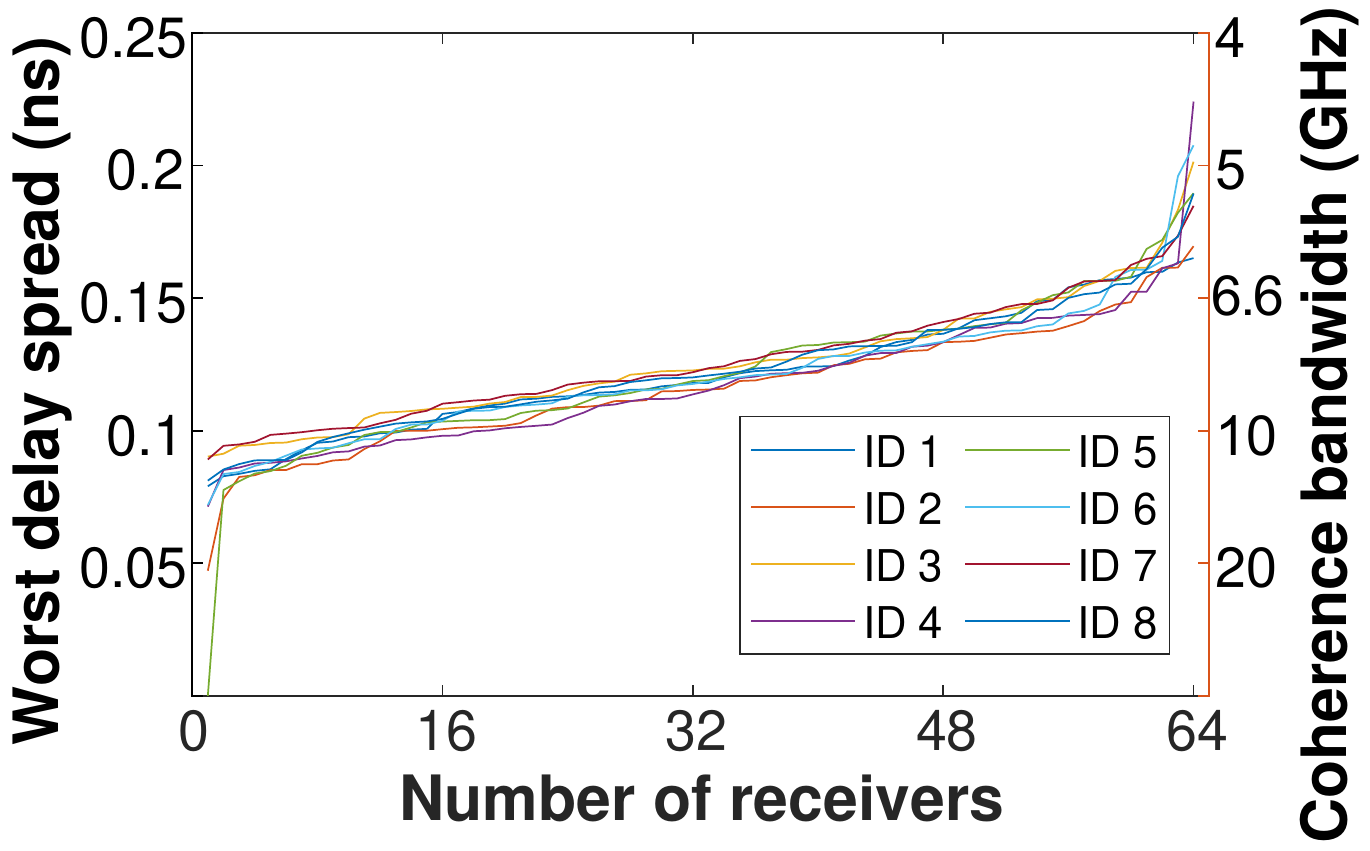}
    \vspace{-0.2cm}
    \caption{Scaling of the delay spread and coherence bandwidth over an increasing number of receivers for the different transmitted symbols in the scenario with three transmitters.}
    \label{fig:worst_ds} 
    \vspace{-0.3cm}
\end{figure}

\begin{figure*}[!t]
    \begin{subfigure}{0.495\textwidth}
    \includegraphics[width=1\textwidth]{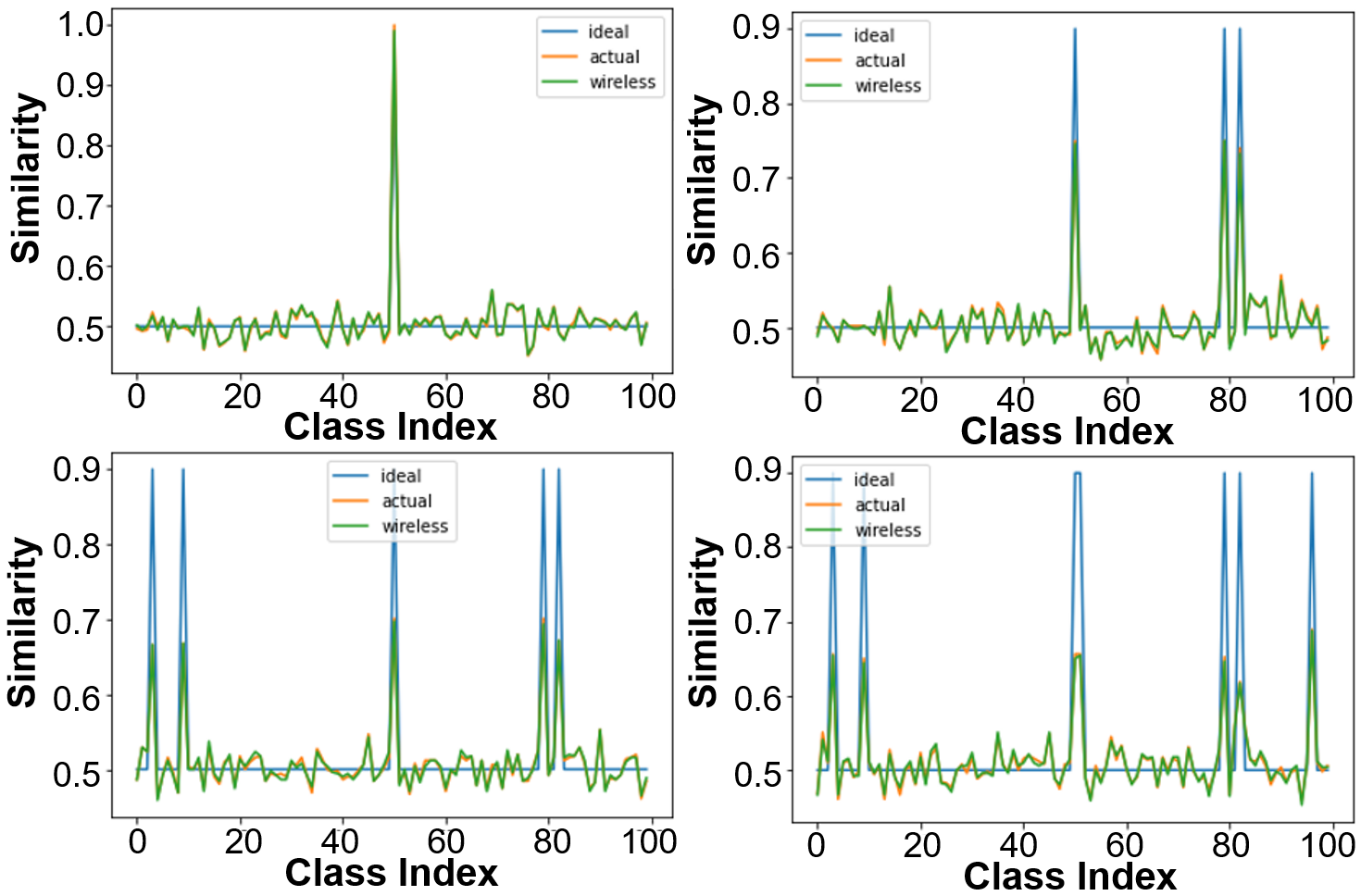}
    \vspace{-0.5cm}
    \caption{Baseline bundling}
    \label{fig:sim_res_baseline} 
    \end{subfigure}
    \hfill
    \begin{subfigure}{0.485\textwidth}
    \includegraphics[width=1\textwidth]{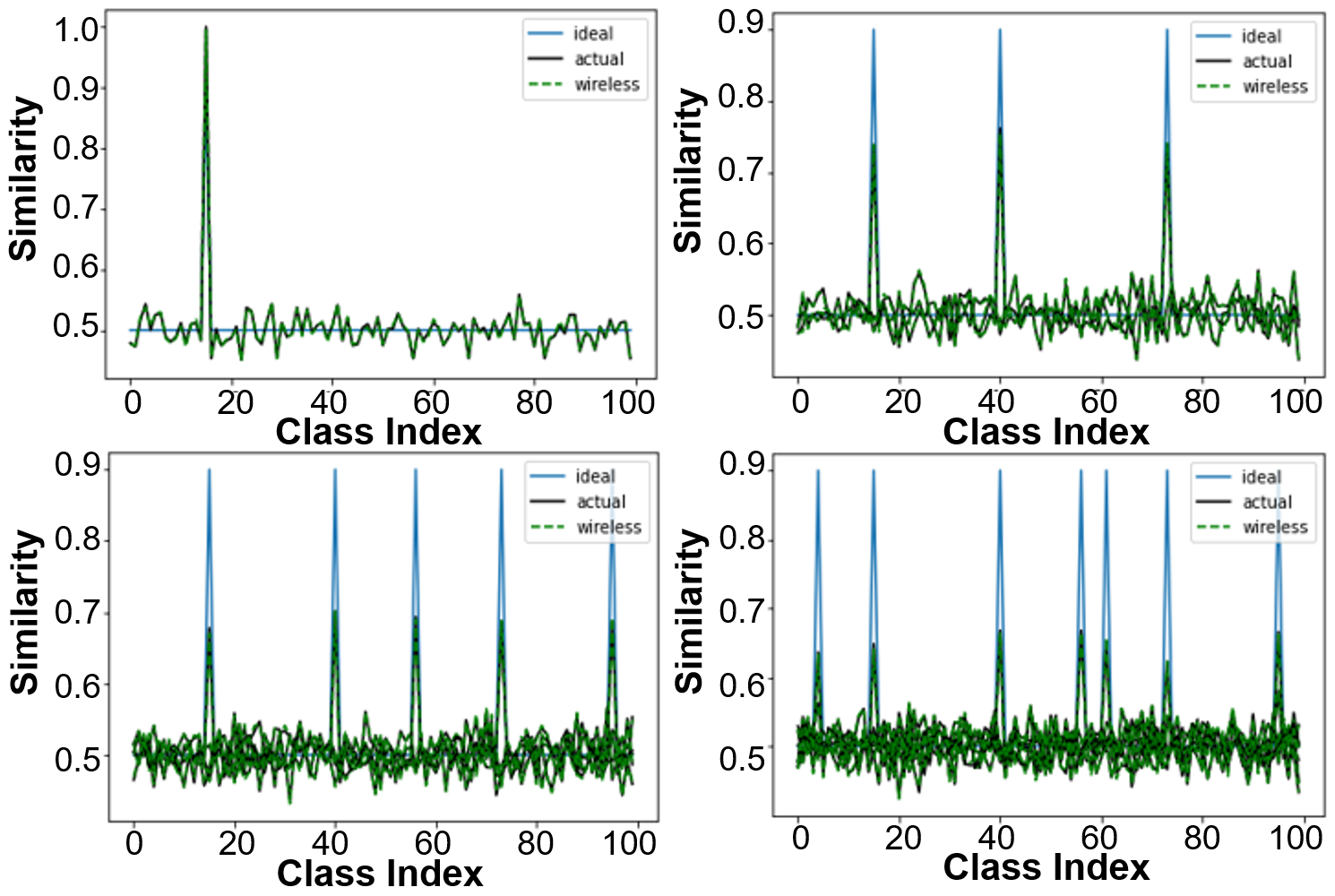}
    \vspace{-0.5cm}
    \caption{Permuted bundling}
    \label{fig:sim_res_permuted} 
    \end{subfigure}
    \vspace{-0.2cm}
    \caption{Similarity results comparison for different forms of bundling one, three, five, and seven hypervectors.}
    \vspace{-0.4cm}
\end{figure*}

To understand how the error rate could scale with the number of receivers, we re-simulate the entire architecture with a varying number of RX cores and computing the average BER obtained in each case. As shown in Fig.~\ref{fig:arch_scal}, the average BER generally increases with the number of receivers for which we are optimizing the architecture. This is expected since, when accommodating more constellations in our optimal TX phases search, we are imposing more conditions and hindering the joint optimization across all receivers.

\vspace{0.1cm} \noindent
\textbf{Time-Domain Analysis.} Still with three transmitters and a number of receivers growing from 1 to 64, we obtain the delay spread and coherence bandwidth of the received signal for each of the $2^3 = 8$ possible transmission combinations. As described earlier, we take the worst-case among all receivers in each simulation. As Fig. \ref{fig:worst_ds} shows, all symbol combinations (marked as different IDs in the figure) have a similar scaling behavior in terms of delay spread. Values lower than 0.1 ns (coherence bandwidth greater than 10 GHz) are obtained consistently for systems with less than 10 receivers. The performance degrades to values around 0.166 ns ($\sim$ 6 GHz) in larger architectures. Given that BPSK has a spectral efficiency of 1 b/s/Hz, the evaluated system would have a total throughput between $3\times 6 = 18$ Gb/s and $3\times 10 = 30$ Gb/s from the encoders to the similarity search engines.

\begin{table}[!t]
\caption{\centering Accuracy in the few-shot learning experiment.}
\vspace{-0.2cm}
\centering
\label{tab:hdframe1}
\begin{tabular}{|c|c|c|c|c|c|c|c|}
\hline 
\multirow{4}{*}{\shortstack{Baseline\\Bundling}} & & \multicolumn{6}{c|}{\textbf{Number of bundled hypervectors}} \\ \cline{3-8}
 & \textbf{Channel} & 1 & 3 & 5 & 7 & 9 & 11\\
\cline{2-8}
 & Ideal & 1 & 0.966 & 0.902 & 0.803 & 0.704 & 0.543\\
\cline{2-8}
 & Wireless & 1 & 0.966 & 0.9 & 0.801 & 0.699 & 0.537 \\ 
\hline \hline
\multirow{4}{*}{\shortstack{Permuted\\Bundling}} & & \multicolumn{6}{c|}{\textbf{Number of bundled hypervectors}} \\ \cline{3-8}
 & \textbf{Channel} & 1 & 3 & 5 & 7 & 9 & 11\\
\cline{2-8}
 & Ideal & 1 & 1 & 1 & 1 & 0.995 & 0.978 \\
\cline{2-8}
 & Wireless & 1 & 1 & 1 & 1 & 0.994 & 0.963\\ 
\hline
\end{tabular} 
\vspace{-0.4cm}
\end{table}

\subsection{Classification Experiments}
\label{sec:HDCeval}
To start the assessment of the HDC-based classification tasks, we first perform the few-shot learning experiment while increasing the BER gradually. As Fig.~\ref{fig:ber_acc} depicts, the class accuracy remains above 99\% even when we apply bit flips equivalent to a BER of 0.26. This means that the noise robustness provided by the HDC properties relaxes the error link conditions, ensuring a correct behaviour under the worst-case wireless scenarios, as we show next.

\vspace{0.1cm} \noindent
\textbf{Few-shot Learning.} Fig.~\ref{fig:sim_res_baseline} and Fig.~\ref{fig:sim_res_permuted} show the similarity search result for the baseline bundling and permuted bundling cases, respectively, in the few-shot learning experiment. The figures show how a single 512-bit query can accommodate several queries via bundling (blue line), and that the wireless system (green line) is able to correctly classify the same queries despite having some bits flipped. 

Table~\ref{tab:hdframe1} shows the numerical results of the final class accuracy for the executed task, comparing an ideal channel without errors with our wireless channel with a sizable BER. The effect of the imperfect bundling is negligible in terms of accuracy, as predicted by Fig.~\ref{fig:ber_acc}. Moreover, the permuted bundling significantly improves the baseline bundling, confirming that the proposed approach supports the aggregation of a dozen hypervectors over the air and the parallelization of similarity search over tens of IMCs.

\begin{figure}[t]
    \centering
    \includegraphics[width=0.85\columnwidth]{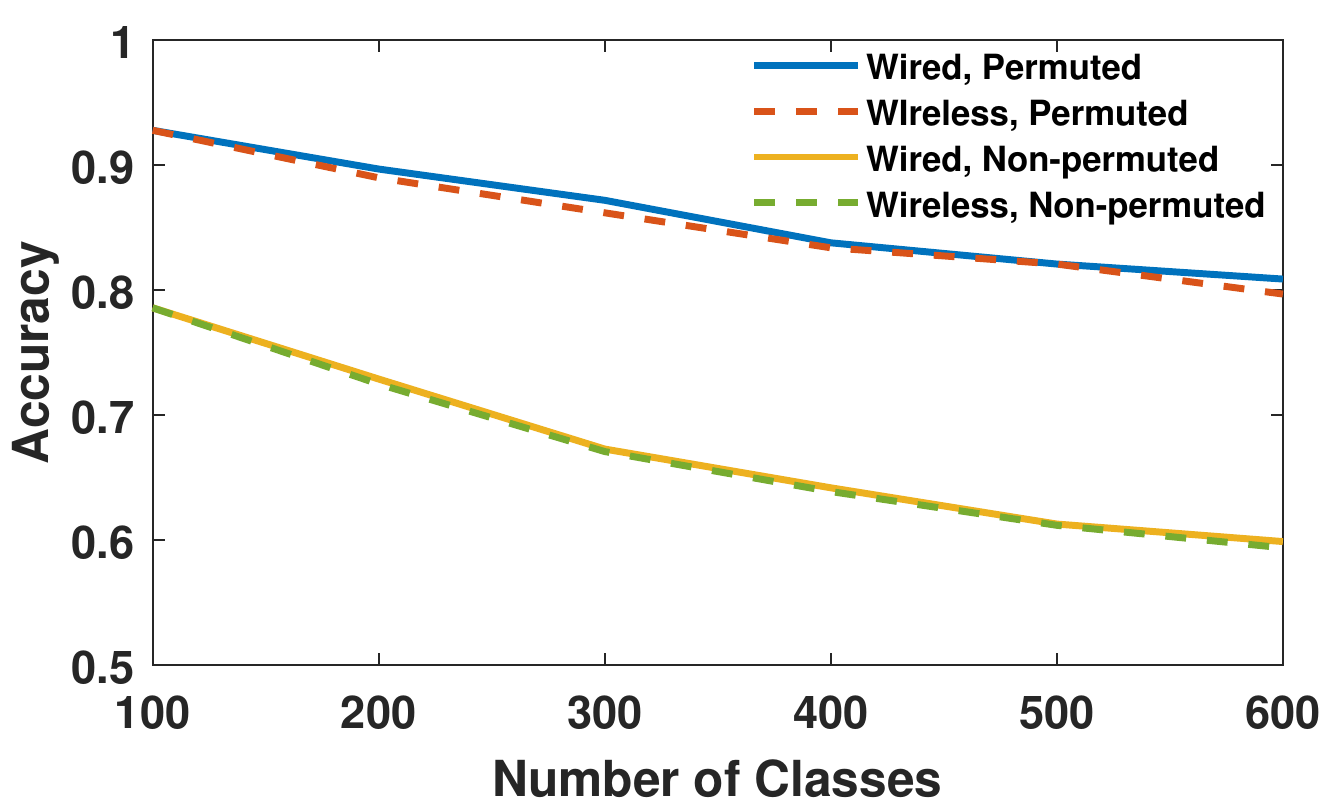}
    \vspace{-0.2cm}
    \caption{Classification accuracy for the continuous learning case with 3 bundled hypervectors.}
    \label{fig:acc_classes3} 
    \vspace{-0.2cm}
\end{figure}

\vspace{0.1cm} \noindent
\textbf{Continual Learning.} Figure \ref{fig:acc_classes3} shows the evolution of the accuracy as the system learns new classes from the initial dictionary of 100 classes, until the entire test set is learnt. Each new set of classes degrades the classification accuracy because possible similarity between classes represented in 512-bit hypervectors. From the figure, it is again clear that (i) the bit flips associated to imperfect OTA bundling have a negligible effect over the classification accuracy, and (ii) permutation provides a consistent improvement over the baseline bundling.

\begin{table}[!t]
\caption{\centering Accuracy in the continual learning experiment.}
\vspace{-0.2cm}
\centering
\label{tab:hdframe1_omniglot}
\begin{tabular}{|c|c|c|c|c|c|c|c|}
\hline 
\multirow{4}{*}{\shortstack{Baseline\\Bundling}} & & \multicolumn{6}{c|}{\textbf{Number of bundled hypervectors}} \\ \cline{3-8}
 & \textbf{Channel} & 1 & 3 & 5 & 7 & 9 & 11\\
\cline{2-8}
 & Ideal & 0.87 & 0.60 & 0.47 & 0.40 & 0.35 & 0.33 \\ 
\cline{2-8}
 & Wireless & 0.81 & 0.59 & 0.47 &  0.39 & 0.35  & 0.32\\ 
\hline \hline
\multirow{4}{*}{\shortstack{Permuted\\Bundling}} & & \multicolumn{6}{c|}{\textbf{Number of bundled hypervectors}} \\ \cline{3-8}
 & \textbf{Channel} & 1 & 3 & 5 & 7 & 9 & 11\\
\cline{2-8}
 & Ideal & 0.87 & 0.81 & 0.73 & 0.63 & 0.55 & 0.46 \\ 
\cline{2-8}
 & Wireless & 0.81 & 0.80 & 0.73 & 0.63 & 0.51 & 0.46 \\ 
\hline
\end{tabular} 
\vspace{-0.4cm}
\end{table}

Table \ref{tab:hdframe1_omniglot} shows the final accuracy for 600 classes and different bundling configurations. The results confirm the conclusions given above in terms of impact of wireless OTA and permuted bundling. It is worth noting that the similarity search over the Omniglot dataset in continual learning does not perform well when bundles exceed a few hypervectors, even with permutations. This suggests that longer hypervectors or an application with a larger dataset would be required to fully benefit from bundling.

\subsection{Comparison with Wired Alternative}
\label{sec:OVHDeval}
Here, we discuss about the performance of WHYPE in terms of speed and the overheads in terms of area and power. To that end, we calculate the area and energy of the entire architecture except for the encoders, which may widely vary depend on the application space, leading to various types of implementations and hardware realizations \cite{joshi2020accurate,HDC_NatElec20,montagna2018pulp}.

To evaluate the area and energy of the similarity search engines, we model them as IMC cores using the design from \cite{HermesVLSI,Y2022khaddamJSSC} as baseline. In more detail, the unit cell at each cross-point of the crossbar consists of 8 transistors and 4 PCM devices (8T4R). The PCM crossbar is back-end-of-the-line integrated with CMOS peripherals, namely, a Pulse-Width Modulation (PWM) circuit which convert a 8-bit digital input vector to an array of time encoded pulses, and an ADC that digitizes the output current of the crossbar to multi-bit values. The area and energy consumption of these components, which are summarized in Table~\ref{tab:areapower2}, have been scaled to consider the technology and dimensions of our architecture. 

For the sake of comparison, we evaluate the area and power of both WHYPE's interconnect and a hypothetical wired interposer-based alternative. In the former case, we consider wireless transmitters and receivers compatible with the requirements of WHYPE \cite{saxena20172,xu201723,multichannel,melamed202230}. In the latter case, the scheme from Figure~\ref{fig:wireless_arch_diagram} is implemented using off-chip interconnects. The majority operation is performed in a central chiplet containing input/output buffers and a 512-wide $M$-input majority gate. The majority and similarity search chiplets are equipped with routers and circuitry to implement serial chip-to-chip links. We assume that the chiplets form a mesh network, similarly to \cite{simba}. Routers are modeled using DSENT \cite{DSENT} with a four-stage pipeline and minimal buffering. Off-chip links are modeled as single-lane links with the energy calculated according to the UCIe standard \cite{UCIe} and the area adapted from \cite{Zimmer2019}. We assume a 32nm technology node with $V_{DD} = 1$ V and $f_{clk} = 1$ GHz. See Table~\ref{tab:areapower} for a summary.

\vspace{0.1cm} \noindent
\textbf{Overhead Analysis.} 
We evaluate the area overhead of the system and the total energy consumed in the hypervector collection, bundling, distribution, and similarity search process using data from Table~\ref{tab:areapower2} and Table~\ref{tab:areapower}. It is observed that the off-chip links and wireless transceivers are the most power-hungry components in the interconnect, while the majority gate circuitry and the routers are also area consuming, mostly due to the wide datapath of 512 bits required to transport the hypervectors. The IMC cores are very efficient in terms of area compared with the interconnect components, while their energy efficiency is penalized by the PWM and ADC circuitry.

\begin{figure}[!t]
    \begin{subfigure}{0.49\columnwidth}
    \includegraphics[width=1\textwidth]{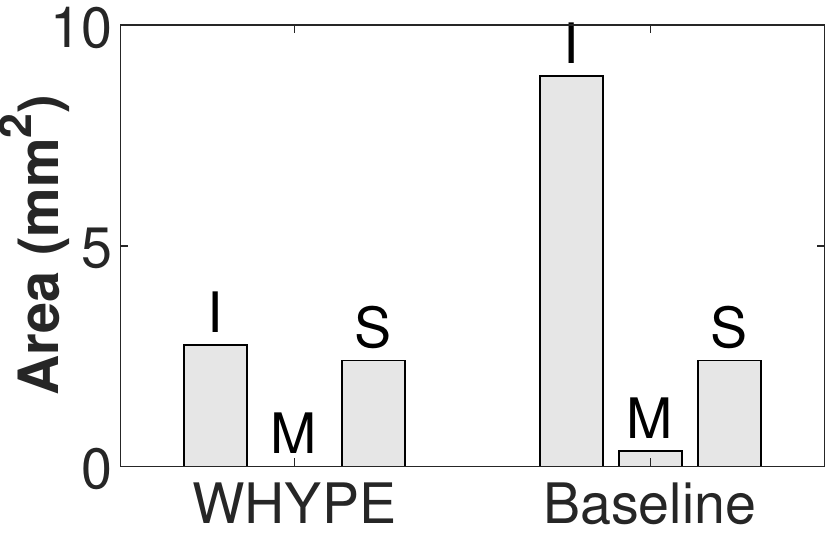}
    \caption{Area}
    \label{fig:areaBreak} 
    \end{subfigure}
    \begin{subfigure}{0.49\columnwidth}
    \includegraphics[width=1\textwidth]{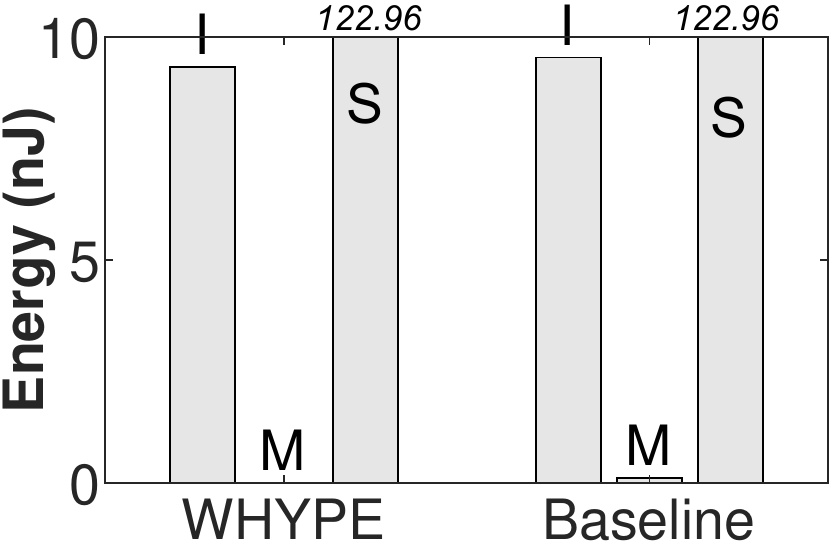}
    \caption{Energy}
    \label{fig:powerBreak} 
    \end{subfigure}
    \vspace{-0.3cm}
    \caption{Area and energy consumption in a system with $M=3$ encoders and $N=8$ search engines, comparing the WHYPE approach with the wired baseline. I, M and S stand for interconnect, majority, and search engines.} \label{fig:breakdown}
    \vspace{-0.2cm}
\end{figure}

\begin{figure}[!t]
    \begin{subfigure}{0.48\columnwidth}
    \includegraphics[width=1\textwidth]{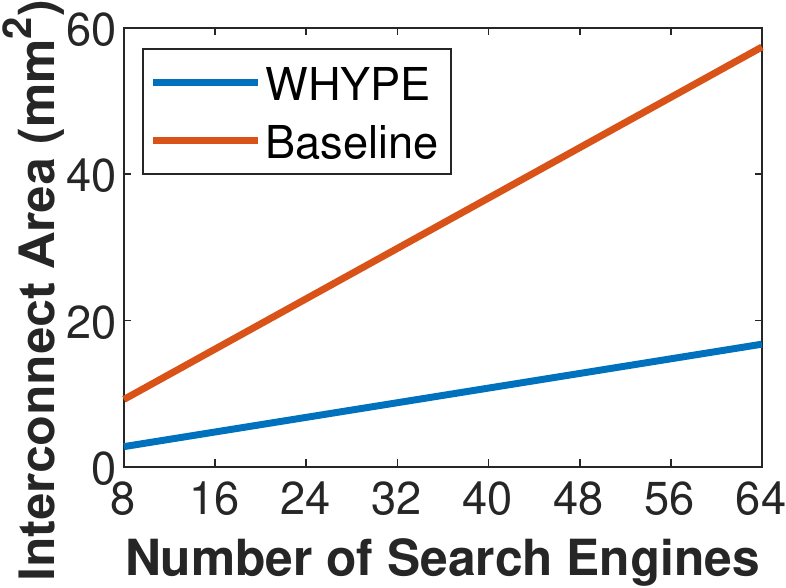}
    \caption{Area overhead}
    \label{fig:area} 
    \end{subfigure}
    \begin{subfigure}{0.5\columnwidth}
    \includegraphics[width=1\textwidth]{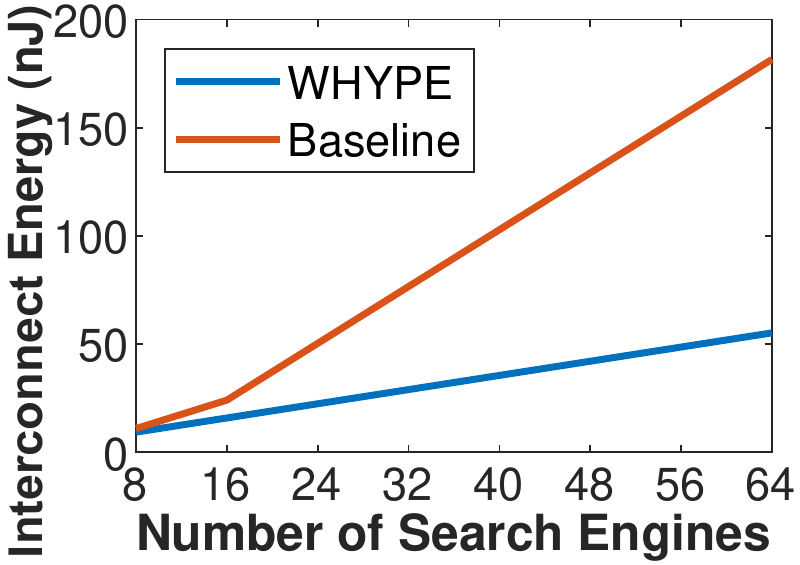}
    \caption{Energy overhead}
    \label{fig:power} 
    \end{subfigure}
    \vspace{-0.2cm}
    \caption{Area and energy overheads of the interconnect with $M=3$ encoders and a variable number $N$ of search engines.} \label{fig:overheads}
    \vspace{-0.2cm}
\end{figure}

Figure \ref{fig:breakdown} shows a breakdown of the area and energy of the HDC system assuming $M=3$ encoders and $N=8$ similarity search engines (512 classes). As can be seen, the interconnect is the most area consuming sub-system, whereas the similarity search is energy consuming due to the large MVM operations performed in the IMC tiles. WHYPE reduces the interconnect area by 3.2$\times$ with a modest effect in the energy because of the small scale of the system. To evaluate scalability, Figure \ref{fig:overheads} shows the area and energy of the interconnect as a function of the number of similarity search engines $N$, for $M=3$. It is observed how, even conservatively assuming single-lane wired links, WHYPE is superior in both area and energy consumption, with a gap that widens as the system is scaled out. This is because WHYPE eliminates any wired connection between chiplets, while the wired alternative needs to traverse the entire system through a mesh topology.

\begin{table}[!t]
\centering
\caption{\centering Area and energy breakdown of IMC for a 512$\times$64 matrix-vector multiplication in 32nm technology.}
\vspace{-0.2cm}
\label{tab:areapower2}
\begin{tabular}{|c|c|c|c|} 
\hline
\multirow{2}{*}{\textbf{Component}}& \textbf{Area} & \textbf{Energy} & \multirow{2}{*}{\textbf{Source}} \\ 
                                   & \textbf{[mm\textsuperscript{2}]} &  \textbf{[nJ]} &    \\  \hline
PCM Crossbar &  0.09    &  0.27   &  \multirow{3}{*}{\shortstack{Scaled from\\ \cite{HermesVLSI,Y2022khaddamJSSC}}} \\ \cline{1-3}
PWM Peripherals & 0.20    &  8.79  &  \\ \cline{1-3}
CCO-based ADC &  0.01     &  6.23  &    \\ \hline
 \end{tabular}
 
PCM = Phase-Change Memory; PWM = Pulse-Width Modulation\\
CCO = Current Controlled Oscillator; ADC = Analog-Digital Converter\\
\vspace{-0.3cm}
\end{table}

\begin{table}[!t]
\centering
\caption{\centering Area and energy breakdown of single instances of interconnect components in 32nm technology.}
\vspace{-0.2cm}
\label{tab:areapower}
\begin{tabular}{c|c|c|c|c|}
\cline{2-5} 
 & \multirow{2}{*}{\textbf{Component}} & \textbf{Area} & \textbf{Energy} & \multirow{2}{*}{\textbf{Source}} \\ 
 &                                  & \textbf{[mm\textsuperscript{2}]} &  \textbf{[pJ/bit]}  & \\  \hline

\multicolumn{1}{|c|}{\multirow{2}{*}{\shortstack{Majority\\Chiplet}}}
              & Majority Gate\S & 0.32  &  0.17 &   \cite{choudhary2019generalized} \\ \cline{2-5} 
\multicolumn{1}{|c|}{}  
            & Buffer & 0.009  & 0.005\ddag & \cite{CACTI} \\ \hline
\multicolumn{1}{|c|}{\multirow{3}{*}{\shortstack{Wired\\Network}}} 
            & Router  & 0.36 &  0.03   &  \multirow{2}{*}{\cite{DSENT}} \\ \cline{2-4} 
\multicolumn{1}{|c|}{}                                             
            & Link (on-chip) & 4$\cdot$10\textsuperscript{-4}  & 0.06    &     \\ \cline{2-5} 
\multicolumn{1}{|c|}{}                                             
            & Link (off-chip)  & 0.25* &  1 & \cite{Zimmer2019,UCIe}  \\ \hline
\multicolumn{1}{|c|}{\multirow{5}{*}{\shortstack{Wireless\\Interface\dag}}} 
            & SerDes  & 0.04  & 0.54  &  \cite{saxena20172}  \\ \cline{2-5} 
\multicolumn{1}{|c|}{}                                             
            & Data Converter & 0.03  & 0.07  &  \cite{xu201723}   \\ \cline{2-5} 
\multicolumn{1}{|c|}{}                                             
            & Transmitter & 0.12 & 1.5  & \cite{multichannel},  \\ \cline{2-4} 
\multicolumn{1}{|c|}{}                                             
            & Receiver & 0.12  & 1.3   & \cite{melamed202230}  \\ \cline{2-5} 
\multicolumn{1}{|c|}{}                                             
            & Antenna  & 0.08  &  N/A  & \cite{gutierrez2009chip} \\ \hline          
\end{tabular} 

Unless noted, the width of the components is 512 bits.\\
\S Five-input, one-output gate.\,\, \ddag Per operation (read, write).\\
\dag Dimensioned to operate at 10 Gb/s.\\ 
*Per pin, serial link operating at 16 Gb/s.
\vspace{-0.1cm}
\end{table}

\vspace{0.1cm} \noindent
\textbf{Bottleneck Analysis.} We assess the performance of WHYPE through an analysis of the latency and throughput of the steps followed since the encoders output the hypervectors until the bundled hypervectors reach the similarity search engines. In the wireless case, the delay analysis is simple: the latency corresponds to the time needed to wirelessly transmit 512 bits. Assuming a 10 Gb/s transmitter, the latency of the majority operation in WHYPE is 51.2 ns independently of the number of encoders or similarity search engines. In contrast, a wired alternative would be bottlenecked by the majority calculation and the communications happening before and after. Latency-wise, the main delay in the wired case is the time required for the hypervectors to travel through the chip-to-chip network. With the conditions here assumed, the delay scales as 2$\sqrt{M+N}$/3 which is the average number of hops to connect two arbitrary chiplets, with each hop taking 36 ns (4 to traverse a router and 32 to transmit 512 bits through a 16 Gb/s serial link). Quickly, the delay becomes much higher than in the wireless case. In comparison, the IMC cores considered here can take between 10 and 128 nanoseconds to realize the similarity calculations, depending on the required accuracy \cite{Y2022khaddamJSSC}. Although majority and similarity operations can be pipelined, a wired NiP would easily bottleneck the system.  

In terms of throughput, WHYPE does not present a bottleneck due to the seamless all-to-all connection between encoders and similarity search engines. At 10 Gb/s of line rate, since all receivers will obtain the bundled hypervector at the same time, the overall throughput is 10$\times$M$\times$N Gb/s. In the hypothetical wired alternative, the bottleneck is the bisection bandwidth of the system. Since the majority chiplet, being in a mesh network, has only four chip-to-chip links, the bisection bandwidth between the encoders and the similarity search engine will be, at most, twice the capacity of the chip-to-chip links. In our assumed scenario, the throughput would be 32 Gb/s. Therefore, WHYPE will be faster even for relatively low values of $M$ and $N$. 

\vspace{0.1cm} \noindent
\textbf{Comparison with 3D Interconnects.} A few works have proposed to implement HDC-based 3D ICs \cite{Li3DVRRAM2016, WuNanotube2018}. One of the reasons could be the reduced link length as compared to the planar NoC or NiP alternatives.  However, the wired nature of 3D ICs and the \emph{AllGather} nature of the communication suggest that the interconnect will continue being a bottleneck. Moreover, 3D ICs can suffer from heat dissipation issues which limit their scaling ability. Due to this, existing efforts only consider up to a couple dozen classes \cite{Li3DVRRAM2016}.

\section{Conclusion} \label{cncl}
In this work, we introduced an OTA on-chip computing concept capable of overcoming the scalability bottleneck present in wired NoC architectures when scaling out IMC-based HDC systems. By using a WNoC communication layer, a number of encoders is able to concurrently broadcast HDC queries towards all the IMC cores within the architecture. Then, a pre-characterization of the propagation environment allows to map the received constellations to the computed composite query, in each core, based on a decision region strategy. Through a proper correspondence between the TX phases, the received constellation and the decision region, we have shown that the opportunistic calculation of the bit-wise majority of the transmitted HDC queries is possible with low error. We demonstrated the concept and shown its scalability up to 11 TXs and 64 RXs, obtaining the BER of the OTA approach and later employing it to evaluate the impact of the WNoC errors in a HDC classification task. Overall, we conclude that the quality of the WNoC links are solid enough to have a negligible impact on the application accuracy, mostly thanks to the great error robustness of HDC.

\section*{Acknowledgment}
Authors gratefully acknowledge funding from the European Union’s Horizon 2020 research and innovation programme under grant agreement No 863337 (WiPLASH), and Horizon Europe research and innovation programme under grant agreement No 101042080 (WINC).

\footnotesize
\bibliographystyle{IEEEtran}
\tiny

\end{document}